\documentclass[lettersize,journal]{IEEEtran}
\usepackage{cite}
\usepackage{amsmath,amssymb,amsfonts}
\usepackage{algpseudocode}
\usepackage{graphicx,color}
\usepackage{textcomp}
\usepackage{algorithm}
\usepackage{cases} 
\usepackage{bm} 
\usepackage{array}
\usepackage[caption=false,font=normalsize,labelfont=sf,textfont=sf]{subfig}
\usepackage{stfloats}
\usepackage{url}
\usepackage{verbatim}
\usepackage{cancel}
\usepackage{multirow}
\usepackage{makecell}

\UseRawInputEncoding
\begin{document}

\title{On Efficient V-BLAST Detection: Recursive and Hybrid Schemes}

\author{Hufei Zhu$^{\text{1}}$, Yanze Zhu$^{\text{2}}$, Qingqing Wu$^{\text{2}}$, Yikui Zhai$^{\text{1}}$, Wen Chen$^{\text{2}}$, and Yang Liu$^{\text{3,4}}$
\thanks{$^{\text{1}}$School of Electronics and Information Engineering, Wuyi University, Jiangmen 529020, Guangdong, China}
\thanks{$^{\text{2}}$School of Information Science and Electronic Engineering, Shanghai Jiao Tong University, Shanghai 200240, China}
\thanks{$^{\text{3}}$School of Information and Communication Engineering, Dalian University of Technology, Dalian 116024, China}
\thanks{$^{\text{4}}$National Mobile Communications Research Laboratory, Southeast University, Nanjing 211189, China}
\thanks{Hufei Zhu and Yanze Zhu contributed equally to this work. CORRESPONDING AUTHOR: Qingqing Wu (e-mail: qingqingwu@sjtu.edu.cn).}
}

\markboth{Journal of \LaTeX\ Class Files,~Vol.~xx, No.~x, 2023}%
{Shell \MakeLowercase{\textit{et al.}}: A Sample Article Using IEEEtran.cls for IEEE Journals}


\maketitle

\begin{abstract}
This paper investigates low-complexity and memory-efficient detection for vertical Bell Laboratories layered space-time architecture (V-BLAST), where recursive and square-root implementations exhibit different tradeoffs in computational complexity. Two new improvements are first developed for the recursive implementation. The first adopts a more efficient partitioned-matrix inverse relation, reducing the dominant complexity of the inverse-construction step by a factor of $1.67$. The second reformulates symbol estimation and interference cancellation using quantities available from the detection-error covariance matrix, thereby avoiding the need to retain the inverse covariance matrix during recursion. The resulting recursive algorithm reduces the dominant complexity of the existing speed-oriented implementation by a factor of $1.3$, while achieving a speedup of approximately $1.86$ and reducing the principal matrix-storage requirement by about one half relative to the existing memory-saving implementation. Since the proposed recursive algorithm offers lower worst-case complexity, whereas the inverse-Cholesky square-root implementation provides lower best-case and average complexities, a hybrid scheme is further developed to switch adaptively between them. The hybrid scheme preserves the worst-case dominant-complexity bound of the recursive implementation, attains the best-case complexity of the square-root implementation, and slightly reduces the average processing cost. Analytical operation counts and numerical results verify these gains and demonstrate robustness to moderate variations in a switching threshold.
\end{abstract}



\maketitle

\section{INTRODUCTION}

Spatial-multiplexing multiple-input multiple-output (MIMO) systems transmit several data streams simultaneously over the same time-frequency resources and can therefore provide substantial gains in spectral efficiency and data rate in rich-scattering environments~\cite{MIMOref9430sd}. Maximum-likelihood (ML) detectors achieve optimum or near-optimum detection performance, but their computational complexity increases rapidly with the number of transmit antennas and the modulation order~\cite{MLreceiver111,MLreceiver222,MLreceiver333}. Vertical Bell Laboratories layered space-time architecture (V-BLAST) offers a more practical performance-complexity tradeoff by combining linear filtering, stream ordering, hard decision, and successive interference cancellation~\cite{zhf1}. In an ordered successive interference cancellation (OSIC) detector, one of the undetected streams is selected at each recursion according to a reliability metric, commonly the post-detection signal-to-noise ratio (SNR). The selected stream is estimated using a zero-forcing or minimum mean-square error (MMSE) filter and quantized to the nearest constellation point. Then its contribution is removed from the received vector. The same procedure is repeated for the reduced-dimensional system until all streams have been detected. Because an incorrect hard decision contaminates all subsequent cancellation stages, the ordering rule plays an important role in limiting error propagation~\cite{BLASTorder4EP}.

A variety of schemes have been developed to mitigate the error-propagation effects of hard-decision OSIC. To improve the performance of conventional SNR-based ordering, the ordering scheme based on the loglikelihood ratio (LLR)~\cite{refZHFLLRorder111, refZHFLLRorder222} exploits the instantaneous noise, while that based on maximum a posteriori probability (MAP)~\cite{refZHFMAPorder} utilizes the symbol decision error. On the other hand, iterative soft interference cancellation (ISIC) reduces the error propagation effects by using soft decisions instead of hard decisions in interference cancellation~\cite{ref7ofDeepSIC,SoftICGlobeCommFoschini,trans10comm348ds32,trans11SP493df4}. For coded systems, ISIC can be integrated with channel decoding through iterative detection and decoding and can approach maximum-likelihood performance with controllable complexity~\cite{ISICnearOptimal93sd,DeepSICEldar}. The ISIC principle is also applicable to direct-sequence code-division multiple-access systems because their multiuser signal model can be expressed in an equivalent MIMO form~\cite{ISICnearOptimal93sd,LMMSEisic2007letter,CDMA1999de34eds3d}. For nonlinear or partially unknown channel models, data-driven implementations have been investigated in which dedicated neural-network modules replace selected model-based components of the conventional detector~\cite{DeepSICEldar}. Neural networks have also been used to detect a subset of the transmitted symbols before cancellation, thereby improving the estimation of the remaining symbols~\cite{refZHFopenJvt}. These works primarily modify the decision rule or the soft-information processing procedure. By contrast, the present study retains the conventional ordered MMSE-OSIC decisions and focuses on reducing the arithmetic and memory costs required to implement them.

Efficient OSIC implementations can be broadly divided into recursive algorithms and square-root algorithms. Recursive methods update the detection-error covariance matrix after each detected stream~\cite{zhf3,zhf5,zhf4,zhf6}. Square-root methods instead propagate the square-root matrix of the detection-error covariance matrix and generally offer improved numerical structure and different ordering-dependent complexity characteristics~\cite{zhf2,refZHFspLetter,refZHFtrans2011,refZHFtrans2015}. Low-complexity implementations have likewise been developed for ISIC detection~\cite{ISICnearOptimalRef11ASIC,transSP2022}. Although all these methods implement closely related linear-estimation and interference-cancellation operations, their arithmetic and memory requirements can differ substantially because they maintain different internal matrix representations.



The OSIC detection has also been incorporated into a wide range of MIMO transmission architectures. Representative applications include linear-dispersion-based perfect space-time codes~\cite{refZHF444}, orthogonal frequency-division multiplexing with index modulation~\cite{refZHF333}, and power-domain nonorthogonal multiple access~\cite{refZHF111,refZHF555,refZHF5b5b5b}. For MIMO filter-bank multicarrier systems, neighborhood-search OSIC detectors based on QR and sorted-QR decompositions were investigated in~\cite{refZHF666}. The OSIC was further combined with ML-based correction and partial tree-search detection for massive-MIMO systems in~\cite{refZHF777} and~\cite{ModOSICs32ds23}, respectively. An exact expression for the average outage probability of zero-forcing OSIC V-BLAST with two transmit antennas and at least two receive antennas was derived in~\cite{refZHF222}. Comprehensive reviews of conventional small- and medium-scale MIMO detection, complexity-scalable detection for large arrays, and massive-MIMO detection can be found in~\cite{refZHFsurvey2}, \cite{refZHFsurvey1}, and~\cite{refZHFsurvey3}, respectively, in which the OSIC detection is consistently discussed.

This paper considers the joint reduction of computational complexity and matrix-storage requirements in recursive V-BLAST detection. The original recursive algorithm in~\cite{zhf3} was progressively improved in~\cite{zhf5} and~\cite{zhf4}, and the resulting techniques were subsequently combined in~\cite{zhf6}. The latter work presented a speed-oriented recursive implementation and a separate memory-saving implementation. The speed-oriented method achieves a lower arithmetic cost but retains both the covariance matrix and its inverse throughout the detection process. In contrast, the memory-saving method avoids storing the inverse covariance matrix, but it cannot exploit several of the arithmetic reductions available to the speed-oriented implementation. Consequently, the two implementations optimize different resources rather than simultaneously minimizing both.

To address this limitation, we first revisit two operations in the speed-oriented recursive algorithm: the construction of the initial detection-error covariance matrix and the recursive interference-cancellation process. For initialization of the covariance matrix, the partitioned-matrix inversion formula~\cite[Ch. 14.12]{ref1BookGlobComm2005} used in~\cite{zhf5} is replaced by an alternative block-inverse relation~\cite[Eq. 8]{InverseSumofMatrix8312}. The resulting update eliminates repeated intermediate operations and reduces the dominant complexity of this stage by a factor of $1.67$. The same block-inverse relation is then used to derive a new interference-cancellation scheme whose recursion depends on the column of the covariance matrix rather than on that of the inverse covariance matrix. This reformulation allows the channel matrix, the inverse covariance matrix, and the covariance matrix to successively reuse the same principal memory region during initialization, while eliminating the need to retain the inverse covariance matrix in the subsequent recursive detection process. The earlier partitioned-matrix inversion can be replaced by the new formulation to propose a recursive algorithm with an additional speed advantage, where the interference cancellation scheme can be replaced by the new one to propose a second implementation with the same dominant arithmetic complexity and a substantially smaller memory requirement. Relative to the existing speed-oriented recursive algorithm, both proposed implementations achieve a dominant-complexity speedup of $1.3$. Relative to the existing memory-saving recursive algorithm, the proposed memory-efficient implementation achieves a speedup of approximately $1.86$ and reduces the principal matrix-storage requirement by about one half.

The square-root implementation based on the inverse Cholesky factor in~\cite{refZHFtrans2011} exhibits a different complexity profile. Compared with the proposed recursive implementation, it has a higher worst-case cost but can achieve lower best-case and average costs, depending on the degree of similarity between the assumed detection order and the actual optimal detection order. This complementary behavior suggests that neither implementation is uniformly preferable over the entire recursion. Then to further reduce the complexity of the proposed recursive implementation, we develop a hybrid implementation, which switches between the proposed memory-efficient recursive implementation and the square-root implementation based on the inverse Cholesky factor by propagating a matrix closely related to the inverse Cholesky factor. The hybrid scheme reduces the best-case complexity of the recursive detector to that of the square-root detector, avoids the unfavorable worst-case complexity of a purely square-root implementation, and slightly lowers the average processing cost of both the recursive and square-root detectors. Furthermore, its worst-case and average complexities do not exceed those of either proposed recursive detector.

The principal contributions of this paper are summarized as follows.

\begin{itemize}

\item The partitioned-matrix inversion lemma utilized in the earlier fastest recursive V-BLAST algorithm is replaced by a computationally efficient block-inverse relation. Accordingly, we reduce the dominant complexity of this stage by a factor of $1.67$, and our corresponding speed-oriented recursive V-BLAST algorithm accelerates the earlier fastest one by a factor of $1.3$.

\item The block-inverse relation is applied to derive a memory-saving interference cancellation scheme, whose recursion uses the covariance matrix instead of the inverse covariance matrix. This enables an in-place recursive implementation in which the channel matrix, the inverse covariance matrix, and the covariance matrix sequentially occupy the same principal matrix-storage region. The proposed interference cancellation scheme is applied in both the proposed speed-oriented recursive detector and the square-root detector based on the inverse Cholesky factor, to propose the corresponding separate memory-saving versions.

\item The proposed memory-saving recursive and square-root detectors require lower worst-case and best-case complexities, respectively. Accordingly, a hybrid detector is proposed to switch adaptively between the proposed recursive and square-root detectors, mainly for reducing the former’s best-case complexity and the latter’s worst-case complexity.

\item The best-case, average, and worst-case dominant complexities are analyzed. Numerical operation counts verify the theoretical results and demonstrate that the hybrid detector is robust to moderate changes in one of its switching thresholds.

\end{itemize}

The remainder of this paper is organized as follows. Section II presents the V-BLAST system model. Then, Section III reviews the covariance-recursive algorithms and the existing complexity-reduction techniques required for the subsequent developments. In Section IV, we derive the proposed recursive implementation with speed advantage and that with both speed advantage and memory saving. In Section V, we introduce the inverse-Cholesky square-root formulation and the hybrid switching scheme. Section VI gives theoretical analysis of the complexities and memory requirements, while Section VII reports the numerical operation counts. Finally, we conclude the paper in Section VIII.


Throughout this paper, uppercase and lowercase boldface letters denote matrices and column vectors, respectively. The superscripts $(\cdot)^{T}$, $(\cdot)^{*}$, $(\cdot)^{H}$, and $(\cdot)^{-1}$ represent transpose, complex conjugate, conjugate transpose, and matrix inverse, respectively. The notation $\mathbf{I}_{M}$ denotes the $M \times M$ identity matrix. MATLAB-style colon notation is used to specify subvectors and submatrices, for example, $\mathbf{z}(1:m)$ denotes the first $m$ entries of $\mathbf{z}$.

\section{SYSTEM MODEL AND V-BLAST DETECTION}
Consider a spatial-multiplexing V-BLAST system with $M$ transmit antennas and $N$ receive antennas, where $N \geq M$. The channel is assumed to be frequency-flat over one detected symbol vector and sufficiently rich in scattering to support simultaneous transmission of $M$ spatial streams. At the transmitter, the input data sequence is demultiplexed into $M$ substreams, which are independently encoded or modulated and transmitted from the corresponding antennas. Denote the vector of transmit symbols from $M$ antennas as
\begin{equation}\label{sMDefine230923}
{\bf{s}}_M=[s_1 ,s_2 , \cdots ,s_M]^T,
\end{equation}
and denote the $N\times M$ complex channel matrix ${\bf{H}}$ as
\begin{equation}\label{equ:10}
{\bf{H}}_M  = [{\bf{h}}_{:1} ,{\bf{h}}_{:2} ,\cdots,{\bf{h}}_{:M}],
\end{equation}
where ${\bf{h}}_{:m}$ ($1 \leq m \leq M$)  is the $m$-th column of ${\bf{H}}$.
Then, the
symbols received by
$N$ antennas
can be written as
\begin{equation}\label{equ:1}
{\bf{x}}^{(M)} = {\bf{H}}_M {\bf{s}}_M + {\bf{n}},
\end{equation}
where ${\bf{n}}$ is the $N\times 1$ complex Gaussian noise vector
with zero mean and covariance $\sigma _n ^2 {\bf{{\rm I}}}_N$. The channel matrix is assumed to be known at the detector, as is standard in the analysis of V-BLAST detection.

The conventional ordered V-BLAST detector processes the $M$ transmitted symbols over $M$ successive recursions. At each recursion, the undetected stream with the highest post-detection SNR is estimated by a linear filter and quantized to the nearest constellation point. Then, its contribution is cancelled from the received signal vector~\cite{zhf1,zhf2,zhf3}. Without loss of generality, we can assume that at the $i$-th ($i = 1, 2, \cdots, M$) recursion, the $m = M - i + 1$ undetected streams and their channel vectors occupy the first $m$ entries of $\mathbf{s}_{M}$ and the first $m$ columns of $\mathbf{H}_{M}$, respectively. The reduced-dimensional model therefore has the same form as (\ref{equ:1}), with $M$ replaced by $m$. The linear MMSE estimate of the remaining vector $\mathbf{s}_{m}$ is then given by
\begin{equation}\label{equ:2}
{\bf{\hat s}}_m = \left( {{\bf{H}}_m^H {\bf{H}}_m + \alpha {\bf{I}}_m}
\right)^{ - 1} {\bf{H}}_m^H {\bf{x}}^{(m)},
\end{equation}
where
$\alpha  \triangleq \frac{\sigma _n^2}{\sigma _s^2}$,
and $\sigma _s^2$ is the power
of each  symbol in ${\bf{s}}_M$.

%

\section{EXISTING RECURSIVE V-BLAST ALGORITHMS}

This section reviews the recursive V-BLAST implementations that form the basis of the subsequent developments. We first describe the original covariance-matrix-based recursion and then summarize four existing improvements designed to reduce its computational or memory requirements. In particular, the matrix quantities and update relations introduced in this section will be repeatedly used in Section IV and in the hybrid implementation developed later.

\subsection{ORIGINAL RECURSIVE V-BLAST ALGORITHM}
The original recursive V-BLAST algorithm developed in~\cite{zhf3} relies on
%
\begin{subnumcases}{\label{RQboth239sd234}}
{\bf{R}}_m={\bf{H}}_{m} ^H {\bf{H}}_{m} + \alpha {\bf{I}}_{m}, &  \label{Rm2Hm3298kds32}\\
{\bf{Q}}_m = {\bf{R}}_m^{ - 1}. &  \label{equ:4}
\end{subnumcases}
%
The matrix ${\bf{Q}}_m$ characterizes the detection-error covariance~\cite{zhf2, zhf3} associated with $ {\bf{e}}_m =
{\bf{s}}_m - {\bf{\hat s}}_m$, and satisfies $E\{{\bf{e}}_m{\bf{e}}_m^H \} =
\sigma _n^2 {\bf{Q}}_m$. Accordingly, among the $m$ undetected symbols at each recursion, the symbol having the highest post detection SNR is identified by the smallest diagonal entry of ${\bf{Q}}_m$.



By applying the Sherman--Morrison formula, the recursive relation in~\cite{zhf3} is obtained as
%
\begin{equation}\label{QinitialOldXiaEqu5}
{{\bf{Q}}_{[n]}} = {{\bf{Q}}_{[n - 1]}} - \frac{{{{\bf{Q}}_{[n - 1]}}{{\bf{h}}_{n:}}{\bf{h}}_{n:}^H{{\bf{Q}}_{[n - 1]}}}}{{1 + {\bf{h}}_{n:}^H{{\bf{Q}}_{[n - 1]}}{{\bf{h}}_{n:}}}},
\end{equation}
%
where ${\bf{h}}_{n:}^H$ denotes the $n$-th row of ${\bf{H}}$. The initial matrices ${\bf{Q}}={\bf{Q}}_{[N]}$ and ${\bf{R}}={\bf{R}}_{[N]}$ are then obtained by successively calculating (\ref{QinitialOldXiaEqu5}) and
\begin{equation}\label{equ_add1}
{\bf{R}}_{[n]}
= \sum\limits_{l =
1}^n {\bf{h}}_{l:}  {\bf{h}}_{l:}^H  + \alpha
{\bf{I}}_M  = {\bf{R}}_{[n - 1]} + {\bf{h}}_{n:}  {\bf{h}}_{n:}^H
\end{equation}
iteratively for $n=1, 2, \cdots, N$, where
${\bf{Q}}_{[0]}  = \frac{1}{\alpha}
{\bf{I}}_M$ and ${\bf{R}}_{[0]}  = \alpha
{\bf{I}}_M$.
For the \textsl{recursion} phase, initialize ${\bf{R}}_{M}={\bf{R}}$, ${\bf{Q}}_{M}={\bf{Q}}$, ${\bf{x}}^{(M)}={\bf{x}}$, and ${\bf{p}} = \left[ 1, 2, \cdots, M \right]^T$.

At the recursion with $m$ ($m=M,M-1,\cdots,2$) undetected symbols, ${\bf{p}}$ is reordered such that the ${p_m}$-th (i.e., ${\bf{p}}(m)$-th) symbol has the highest SNR among the remaining symbols. Then, ${\bf{H}}_{m}$, ${\bf{R}}_{m}$, and ${\bf{Q}}_{m}$ are permuted accordingly to ensure that the ${p_m}$-th symbol can be estimated as
%
\begin{equation}\label{ypm2qmHmxm23923d3}
\hat s_{{p_m}} = {\bf{q}}_m^H {\bf{H}}_{m}^H {\bf{x}}^{(m)},
\end{equation}
where ${\bf{q}}_m$ denotes the $m$-th column of ${\bf{Q}}_{m}$. The estimate $\hat s_{{p_m}}$ is then quantized to $s_{{p_m}}$ according to the constellation in use. To proceed to the next recursion, the interference of $s_{{p_m}}$ is removed from ${\bf{x}}^{(m)}$, which yields
%
\begin{equation}\label{equ:BLAST_ICorig}
{\bf{x}}^{(m-1)}  =
{\bf{x}}^{(m)}  -  s_{{p_m}} {\bf{h}}_{:{{m}}},
\end{equation}
whereas ${\bf{R}}_{m-1}$ is identified from the following partition of
%
\begin{equation}\label{equ:11}
{\bf{R}}_{m}  = \left[ {\begin{array}{*{20}c}
{{\bf{R}}_{m-1} } & {{\bf{v}}_{m-1}}  \\
{{\bf{v}}_{m-1}^H } & {\gamma _{m} }  \\
\end{array}} \right].
\end{equation}
%
Meanwhile, ${\bf{Q}}_{m}$ is deflated to ${\bf{Q}}_{m - 1}$ according to

\begin{equation}\label{QdeflatOriginal3sd32}
{\bf{Q}}_{m - 1}^{} = {\bf{\bar Q}}_{m - 1}^{} - \frac{{{\bf{\bar Q}}_{m - 1}^{}{\bf{v}}_{m-1}^{}{\bf{v}}_{m-1}^H{\bf{\bar Q}}_{m - 1}^{}}}{{{\gamma _m} + {\bf{v}}_{m-1}^H{\bf{\bar Q}}_{m - 1}^{}{\bf{v}}_{m-1}^{}}},
\end{equation}

where ${\bf{\bar Q}}_{m-1}$ is identified from the following partition~\cite[Eq. (13)]{zhf6} of

\begin{equation}\label{equQsmally26}
{\bf{Q}}_m  = \left[ {\begin{array}{*{20}c}
{{\bf{\bar Q}}_{m - 1}^{} } & {{\bf{w}}_{m-1}^{} }  \\
{{\bf{w}}_{m-1}^H } & {\omega _{m} }  \\
\end{array}} \right].
\end{equation}

In (\ref{equ:11})--(\ref{equQsmally26}), ${\bf{v}}_{m-1}$ and ${{\bf{w}}_{m-1}^{}}$ are obtained by removing the last entries of ${\bf{r}}_m$ and ${\bf{q}}_m$, respectively, where ${\bf{r}}_m$ and ${\bf{q}}_m$ denote the $m$-th columns of ${\bf{R}}_{m}$ and ${\bf{Q}}_{m}$.

\subsection{EXISTING EFFICIENT RECURSIVE V-BLAST ALGORITHMS}
In the original recursive algorithm, the dominant computational complexity is of the order of $O({M^3}+{M^2}N)$. This complexity mainly arises from the initialization of ${\bf{Q}}$ and ${\bf{R}}$, the estimation of $s_{p_m}$, and the deflation of ${\bf{Q}}_m$ for $2 \le m \le M$, which are performed by (\ref{QinitialOldXiaEqu5}), (\ref{equ_add1}), (\ref{ypm2qmHmxm23923d3}), and (\ref{QdeflatOriginal3sd32}), respectively. To reduce the computational burden of the original recursive algorithm, Improvements \uppercase\expandafter{\romannumeral+1}--\uppercase\expandafter{\romannumeral+4} were successively proposed in \cite{zhf5}, \cite{zhf4} and \cite{zhf6}.

\noindent {\bf{Improvement \uppercase\expandafter{\romannumeral+1}:}}
To accelerate the computation of the initial ${\bf{Q}}$, the Sherman--Morrison update in (\ref{QinitialOldXiaEqu5}), as adopted in \cite{zhf3}, is replaced in \cite{zhf5} by the partitioned-matrix inversion lemma~\cite[Ch. 14.12]{ref1BookGlobComm2005}. Specifically, for ${\bf{R}}_{m}$ partitioned as in (\ref{equ:11}) and its inverse ${\bf{Q}}_{m}$ partitioned as in (\ref{equQsmally26}), ${\bf{\bar Q}}_{m-1}$, ${\bf{w}}_{m-1}$, and $\omega_{m}$ are obtained from

\begin{subnumcases}{\label{derive2Globecomm2005all3}}
{\bf{\bar Q}}_{m-1}^{}  = {\bf{Q}}_{m-1}^{}  + \frac{{{\bf{Q}}_{m-1}^{}
{\bf{v}}_{m-1}^{} {\bf{v}}_{m-1}^H {\bf{Q}}_{m-1}^{} }}{{\gamma _{m}
- {\bf{v}}_{m-1}^H {\bf{Q}}_{m-1}^{} {\bf{v}}_{m-1}^{} }},
&  \label{equ:13OldJul14}\\
{\bf{w}}_{m-1}^{}  =   { - \gamma _{m}^{ - 1} {\bf{\bar Q}}_{m-1} {\bf{v}}_{m-1} },
& \label{derive2aGlobecomm2005eq2}\\
\omega _{m}  = {\gamma _{m}^{ - 1}
+ \gamma _{m}^{ - 2} {\bf{v}}_{m-1}^H {\bf{\bar Q}}_{m-1}^{} {\bf{v}}_{m-1} }.   & \label{derive2cGlobecomm2005eq3}
\end{subnumcases}

In \cite{zhf5}, (\ref{equQsmally26}) and (\ref{derive2Globecomm2005all3}) are applied to construct ${\bf{Q}}_m={\bf{R}}_m^{-1}$ from ${\bf{Q}}_{m-1}$ for $m=2,3,\ldots,M$. Starting from

\begin{equation}\label{Q1compute298jd32}
{\bf{Q}}_1 =1/{\bf{R}}_1,
\end{equation}

this recursion yields the initial matrix ${\bf{Q}}={\bf{Q}}_M$.

\noindent {\bf{Improvement \uppercase\expandafter{\romannumeral+2}:}}
In \cite{zhf4}, the symbol-estimation expression in (\ref{ypm2qmHmxm23923d3}) is reformulated as

\begin{equation}\label{equ:16}
\hat s_{p_m }  = {\bf{q}}_m^H  {\bf{z}}_m,
\end{equation}

where ${\bf{z}}_m$ is defined by

\begin{equation}\label{zm2HmxmDef3290sd23}
{\bf{z}}_m  = {\bf{H}}_m^H {\bf{x}}^{(m)}.
\end{equation}

Only the initial vector ${\bf{z}}_M$ is directly computed from (\ref{zm2HmxmDef3290sd23}) by setting $m=M$. Then, ${\bf{z}}_{m-1}$ ($m=M,M-1,\cdots,2$) is efficiently obtained by removing the contribution of $s_{p_m}$ from the permuted ${\bf{z}}_m$ through the recursion

\begin{equation}\label{equ:BLAST_IC}
{\bf{z}}_{m - 1}  =
{\bf{\bar z}}_m  -  s_{p_m }
{{\bf{v}}_{m-1}},
\end{equation}

where ${\bf{\bar z}}_m$ is obtained by removing the last entry of ${\bf{z}}_m$.

\noindent {\bf{Improvement \uppercase\expandafter{\romannumeral+3}:}}
In \cite[Algorithm II]{zhf5} and \cite{zhf4}, the Hermitian structure of the involved matrices is exploited to simplify the initialization and deflation of ${\bf{Q}}$ using (\ref{QinitialOldXiaEqu5}) and (\ref{QdeflatOriginal3sd32}), respectively.

\noindent {\bf{Improvement \uppercase\expandafter{\romannumeral+4}:}}
In \cite{zhf6}, the deflation of ${\bf{Q}}_m$ is accelerated by replacing (\ref{QdeflatOriginal3sd32}) with

\begin{equation}\label{equ:14}
{\bf{Q}}_{m - 1}  = {\bf{\bar Q}}_{m - 1}  - \omega _{m}^{ - 1}
{\bf{w}}_{m-1}^{} {\bf{w}}_{m-1}^H,
\end{equation}

where ${\bf{\bar Q}}_{m - 1}$, ${\bf{w}}_{m-1}$, and $\omega_{m}$ are the corresponding components of ${\bf{Q}}_{m}$, as specified in (\ref{equQsmally26}).




In \cite{zhf6}, Improvements \uppercase\expandafter{\romannumeral+1}--\uppercase\expandafter{\romannumeral+3} were first incorporated into the original algorithm, resulting in the ``fastest known algorithm" prior to \cite{zhf6}. Improvement \uppercase\expandafter{\romannumeral+4} was subsequently introduced to further enhance this algorithm, leading to the algorithm with speed advantage summarized in \textbf{Algorithm 1}.

\begin{algorithm}[!t]
\caption{Recursive V-BLAST Algorithm with Speed Advantage in \cite{zhf6}}
\renewcommand{\algorithmicrequire}{\textbf{Initialization:}} 
\renewcommand{\algorithmicensure}{\textbf{Recursion:}}
\begin{algorithmic}[1]
\Require Set
${\bf{p}} = \left[ 1, 2, \cdots,
M \right]^T$ and compute $ {\bf{z}}_M  = {\bf{H}}^H {\bf{x}}$;  
Compute (\ref{equ_add1}) iteratively for $n=1, 2, \cdots, N$
to obtain the initial ${\bf{R}}_M={\bf{R}}_{[N]}$; 
Compute  ${\bf{Q}}_1$ by
(\ref{Q1compute298jd32}), and then compute
(\ref{equQsmally26})
and  (\ref{derive2Globecomm2005all3}) iteratively
for $m=2, 3, \cdots, M$, to obtain  the initial $ {\bf{Q}}_M$;
\Ensure For $m=M,M-1,\cdots,2$:
\State  Find $ l_m  = \mathop {\arg \min }\limits_{j = 1,2...}^m {q}_{jj} $, where
$q_{jj}$
is the $j$-th diagonal entry of  ${\bf{Q}}_{m}$;
Permute entries $l_m$ and $m$ in ${\bf{p}}$ and ${\bf{z}}_m$;
Permute rows and columns $l_m$
and $m$ in ${\bf{R}}_{m}$ and ${\bf{Q}}_{m}$;
\State  Compute $\hat s_{{p_m}}$ by (\ref{equ:16}),
which is quantized
to  $s_{{p_m}}$;
\State Cancel the effect of $s_{{p_m}}$ in ${\bf{z}}_m$ to obtain ${\bf{z}}_{m-1}$ by
(\ref{equ:BLAST_IC}),
and deflate
${{\bf{Q}}_{m}}$ to ${{\bf{Q}}_{m-1}}$ by (\ref{equ:14});
\renewcommand{\algorithmicrequire}{\textbf{Solution:}} 
\Require When $m = 1$, only run step 2 to get $s_{p_1}$.
\end{algorithmic}
\end{algorithm}

On the other hand, the memory-saving algorithm proposed in \cite{zhf6} avoids computing ${\bf{R}}$, thereby eliminating the storage and permutation overhead associated with this matrix. It incorporates only Improvements \uppercase\expandafter{\romannumeral+3} and \uppercase\expandafter{\romannumeral+4} into the original algorithm, while Improvements \uppercase\expandafter{\romannumeral+1} and \uppercase\expandafter{\romannumeral+2} are not adopted because they require entries of ${\bf{R}}$. Furthermore, after incorporating Improvement \uppercase\expandafter{\romannumeral+4}, the deflation of ${\bf{Q}}_m$ in the \textsl{recursion} phase relies on entries of ${\bf{Q}}$ rather than those of ${\bf{R}}$. Consequently, $ {\bf{R}}_M$ no longer needs to be computed using (\ref{equ_add1}) during the \textsl{initialization} phase. The memory-saving algorithm in \cite{zhf6} is summarized in \textbf{Algorithm 2}.

\begin{algorithm}[!t]
	\caption{Recursive V-BLAST Algorithm with Memory Saving in \cite{zhf6}}
	\renewcommand{\algorithmicrequire}{\textbf{Initialization:}} 
	\renewcommand{\algorithmicensure}{\textbf{Recursion:}}
	\label{alg:Framwork}
	\begin{algorithmic}[1] %
		\Require Set
		${\bf{p}} = \left[ 1, 2, \cdots,
		M \right]^T$ and ${\bf{H}}_{M}={\bf{H}}$;  
		Compute
		(\ref{QinitialOldXiaEqu5})  iteratively for $n=1, 2, \cdots, N$
		to obtain the initial ${\bf{Q}}_M={\bf{Q}}_{[N]}$;
		\Ensure For $m=M,M-1,\cdots,2$:
		\State Find $ l_m  = \mathop {\arg \min }\limits_{j = 1,2...}^m {q}_{jj} $, where
		$q_{jj}$
		is the $j$-th diagonal entry of ${\bf{Q}}_{m}$;
		Permute entries $l_m$ and $m$ in ${\bf{p}}$;  Permute columns $l_m$
		and $m$ in ${\bf{H}}_{m}$;
		Permute rows and columns $l_m$
		and $m$ in ${\bf{Q}}_{m}$;
		\State Compute $\hat s_{{p_m}}$ by  (\ref{ypm2qmHmxm23923d3}),
		which is quantized
		to  $s_{{p_m}}$;
		\State   Cancel the effect of $s_{{p_m}}$ in ${\bf{x}}^{(m)}$ to obtain ${\bf{x}}^{(m-1)}$ by (\ref{equ:BLAST_ICorig});
		Deflate
		${{\bf{Q}}_{m}}$ to ${{\bf{Q}}_{m-1}}$ by (\ref{equ:14}); Remove the last column of ${\bf{H}}_{m}$ to obtain ${\bf{H}}_{m-1}$;
		\renewcommand{\algorithmicrequire}{\textbf{Solution:}} 
		\Require When $m = 1$, only run step 2 to get $s_{p_1}$.
	\end{algorithmic}
\end{algorithm}

To the best of our knowledge, among the existing recursive V-BLAST algorithms in \cite{zhf3,zhf4,zhf5,zhf6}, the algorithm with speed advantage (i.e., \textbf{Algorithm 1}) and that with memory saving (i.e., \textbf{Algorithm 2}) in \cite{zhf6} achieve the lowest computational complexity and memory requirement, respectively.

		
\section{PROPOSED RECURSIVE V-BLAST ALGORITHM}

In this section, Improvements \uppercase\expandafter{\romannumeral+5} and \uppercase\expandafter{\romannumeral+6} are proposed to replace Improvements \uppercase\expandafter{\romannumeral+1} and \uppercase\expandafter{\romannumeral+2}, respectively. Improvement \uppercase\expandafter{\romannumeral+5} reduces the computational cost of the partitioned-matrix inversion employed in Improvement \uppercase\expandafter{\romannumeral+1}. Based on the formulas developed for Improvement \uppercase\expandafter{\romannumeral+5}, Improvement \uppercase\expandafter{\romannumeral+6} is further derived. Compared with Improvement \uppercase\expandafter{\romannumeral+2}, it reduces the memory requirement without sacrificing computational speed by using ${\bf{Q}}$ rather than ${\bf{R}}$ for interference cancellation during the \textsl{recursion} phase.


\subsection{IMPROVEMENT \uppercase\expandafter{\romannumeral+5}: PARTITIONED-MATRIX INVERSION WITH LOW-COMPLEXITY}
Instead of employing the partitioned-matrix inversion lemma~\cite[Ch. 14.12]{ref1BookGlobComm2005} adopted in \cite{zhf5}, the proposed Improvement \uppercase\expandafter{\romannumeral+5} applies the partitioned-matrix inversion identity~\cite[Eq. 8]{InverseSumofMatrix8312}\footnote{The partitioned-matrix inversion identity was not originally proposed in \cite{InverseSumofMatrix8312}. Several studies published between 1917 and 1978 were cited near equation (8) therein to discuss its first explicit presentation.} to obtain the inverse of ${\bf{R}}_m$. Specifically, for ${\bf{R}}_m$ partitioned as in (\ref{equ:11}), its inverse ${\bf{Q}}_m$ is partitioned as in (\ref{equQsmally26}), where

\begin{subnumcases}{\label{derive2}}
\omega _{m}  = \left( {\gamma _{m}  - {\bf{v}}_{m-1}^H
{\bf{Q}}_{m-1}^{}
{\bf{v}}_{m-1}^{} } \right)^{ - 1}, &  \label{derive2b}\\
{\bf{w}}_{m-1}^{}  =  - \omega _{m} {\bf{Q}}_{m-1}^{} {\bf{v}}_{m-1}^{}, & \label{derive2a}\\
{\bf{\bar Q}}_{m-1}^{}  = {\bf{Q}}_{m-1}^{}  + \omega _{m}^{ - 1}
{\bf{w}}_{m-1}^{} {\bf{w}}_{m-1}^H. & \label{derive2c}
\end{subnumcases}

The above equation (\ref{derive2}) is derived from equation (8) in \cite{InverseSumofMatrix8312}, which provides a substantially simpler formulation than (\ref{derive2Globecomm2005all3}) used in Improvement \uppercase\expandafter{\romannumeral+1}. The derivations of (\ref{derive2}) are relegated to Appendix A. To eliminate the unnecessary division required to calculate $\omega_{m}^{ - 1}$ in (\ref{derive2c}), (\ref{derive2}) is further rewritten as

\begin{subnumcases}{\label{derive3}}
{\bf{\tilde w}}_{m-1}  = {\bf{Q}}_{m-1}^{} {\bf{v}}_{m-1}, & \label{get_u_less_div}\\
\omega _{m}  = \left( {\gamma _{m}  - {\bf{v}}_{m-1}^H {\bf{\tilde w}}_{m-1} }
\right)^{ - 1}, & \label{derive3b}\\
{\bf{w}}_{m-1}^{}  =  - \omega _{m} {\bf{\tilde w}}_{m-1}, &  \label{derive3a}\\
{\bf{\bar Q}}_{m-1}^{}  = {\bf{Q}}_{m-1}^{}  + \omega _{m}^{}
{\bf{\tilde w}}_{m-1}^{} {\bf{\tilde w}}_{m-1}^H. & \label{derive3c}
\end{subnumcases}

In \textbf{Algorithm 1}, i.e., the improved algorithm with speed advantage in \cite{zhf6}, the initial ${\bf{Q}}$ is computed using (\ref{derive3}) from Improvement \uppercase\expandafter{\romannumeral+5}, rather than (\ref{derive2Globecomm2005all3}) from Improvement \uppercase\expandafter{\romannumeral+1}. This replacement yields the proposed recursive algorithm with speed advantage.

\subsection{IMPROVEMENT \uppercase\expandafter{\romannumeral+6}: INTERFERENCE CANCELLATION WITH MEMORY-SAVING}

In the proposed Improvement \uppercase\expandafter{\romannumeral+6}, (\ref{derive3}) developed for Improvement \uppercase\expandafter{\romannumeral+5} is used to derive an improved memory-saving interference-cancellation scheme. This scheme eliminates the use of ${\bf{R}}_M$ during the \textsl{recursion} phase, thereby avoiding the memory required to store ${\bf{R}}_M$ and the corresponding permutation operations. For ease of exposition, the detection order in this subsection is assumed to be $M, M-1, \cdots, 1$.


To eliminate the dependence on ${{\bf{v}}_{m-1}}$ from ${\bf{R}}_m$ when updating ${\bf{z}}_m$ to ${\bf{z}}_{m-1}$ through (\ref{equ:BLAST_IC}), the initial vector ${\bf{z}}_M$ is kept unchanged throughout the \textsl{recursion} phase. Based on this invariant vector, define

\begin{equation}\label{defd329023dks32d3}
{{\bf{d}}_m} = {{\bf{Q}}_m}\left({\bf{z}}_M(1:m) - {{\bf{z}}_m}\right),
\end{equation}

where ${\bf{z}}_M(1:m)$ denotes the first $m$ entries of ${\bf{z}}_M$. When $m=M$, (\ref{defd329023dks32d3}) gives the initialization

\begin{equation}\label{d20wesddswe3}
{{\bf{d}}_{M}}={{\bf{0}}_{M}}.
\end{equation}

The vector ${{\bf{d}}_m}$ is then used to estimate $s_{p_m}$ as

\begin{equation}\label{equ:16Update}
\hat s_{p_m }  = {\bf{q}}_m^H{\bf{z}}_M(1:m) - {{\bf{d}}_m}(m),
\end{equation}

and can be recursively updated to ${{\bf{d}}_{m - 1}}$ according to

\begin{equation}\label{dUpdate2388932sde3}
{{\bf{d}}_{m - 1}} = {\bf{\bar d}}_m  - \left( {{{s}_{{p_m}}}+{\bf{d}}_m^{}(m)} \right){\bf{w}}_{m-1}/\omega _{m},
\end{equation}

where ${{\bf{d}}_m}(m)$ is the last entry of ${{\bf{d}}_m}$, and ${\bf{\bar d}}_m$ is formed by removing this entry, i.e.,

\begin{equation}\label{d2twoParts2390sd234}
{\bf{d}}_m = [{\bf{\bar d}}_m^T, {{\bf{d}}_m}(m)]^T.
\end{equation}


To derive (\ref{equ:16Update}), simply rearrange (\ref{defd329023dks32d3}) as

\begin{equation}\label{QlastEntry3298382dse}
{\bf{q}}_m^H{{\bf{z}}_m} = {\bf{q}}_m^H{\bf{z}}_M(1:m) - {{\bf{d}}_m}(m),
\end{equation}

and then substitute (\ref{QlastEntry3298382dse}) into (\ref{equ:16}). The recursion in (\ref{dUpdate2388932sde3}) will be derived in the remainder of this subsection. It removes the interference of $s_{p_m}$ from ${\bf{d}}_m$, thereby enabling an efficient update from ${\bf{d}}_m$ to ${\bf{d}}_{m-1}$.

First, we verify that ${\bf{\bar d}}_m$, ${\bf{d}}_m^{}(m)$, and ${{\bf{d}}_{m - 1}}$ in (\ref{dUpdate2388932sde3}) satisfy
\begin{subnumcases}{\label{all3equ2Deduce26}}
{\bf{\bar d}}_m = [{\bf{Q}}_{m - 1}^{} + \omega _{m}^{}
{\bf{\tilde w}}_{m-1}^{} {\bf{\tilde w}}_{m-1}^H, - \omega _{m} {\bf{\tilde w}}_{m-1}] \notag\\ \qquad\times \left( {{\bf{z}}_M(1:m) - {{\bf{z}}_m}} \right), & \label{dminus1dse32sd32before}\\
{\bf{d}}_m^{}(m) = [- \omega _{m} {\bf{\tilde w}}_{m-1}^H, {{\omega _m}}]\left( {{\bf{z}}_M(1:m) - {{\bf{z}}_m}} \right), & \label{d2wpsizz239ds32d} \\
{{\bf{d}}_{m - 1}} \!=\! {{\bf{Q}}_{m - 1}}\left( {{\bf{z}}_M(1\!:\!m\!-\!1) \!-\! {\bf{\bar z}}_m} \right)\!+\!{{s}_{{p_m}}}{\bf{\tilde w}}_{m-1}.  &  \label{d2Qzzsv392dsk3}
\end{subnumcases}
Substituting (\ref{derive3a}) and (\ref{derive3c}) into (\ref{equQsmally26}) gives

\begin{equation}\label{equ:15T2Q3832}
{\bf{Q}}_m  = \left[ {\begin{array}{*{20}c}
{\bf{Q}}_{m-1}^{}  + \omega _{m}^{}
{\bf{\tilde w}}_{m-1}^{} {\bf{\tilde w}}_{m-1}^H & - \omega _{m} {\bf{\tilde w}}_{m-1}  \\
- \omega _{m} {\bf{\tilde w}}_{m-1}^H   & {\omega _{m} }  \\
\end{array}} \right].
\end{equation}

Equations (\ref{dminus1dse32sd32before}) and (\ref{d2wpsizz239ds32d}) then follow by substituting (\ref{equ:15T2Q3832}) and (\ref{d2twoParts2390sd234}) into (\ref{defd329023dks32d3}). To verify (\ref{d2Qzzsv392dsk3}), substitute (\ref{equ:BLAST_IC}) into (\ref{defd329023dks32d3}) with $m=m-1$ to yield

\begin{align}
{{\bf{d}}_{m - 1}} &= {{\bf{Q}}_{m - 1}}\left( {{\bf{z}}_M(1:m-1) - {\bf{\bar z}}_m + {{s}_{{p_m}}}{{\bf{v}}_{m-1}}} \right)  \notag  \\
&=  {{\bf{Q}}_{m - 1}}\left( {{\bf{z}}_M(1:m-1) - {\bf{\bar z}}_m} \right)+{{s}_{{p_m}}}{{\bf{Q}}_{m - 1}} {{\bf{v}}_{m-1}}, \notag
\end{align}

and (\ref{d2Qzzsv392dsk3}) is obtained by further applying (\ref{get_u_less_div}).
Equation (\ref{all3equ2Deduce26}) is now used to derive (\ref{dUpdate2388932sde3}). Substituting (\ref{dminus1dse32sd32before}) and (\ref{d2Qzzsv392dsk3}) into ${{\bf{d}}_{m - 1}}-{\bf{\bar d}}_m$ gives

\begin{small}
	\begin{multline*}\label{}
		{{\bf{d}}_{m - 1}} -  {\bf{\bar d}}_m  = {{\bf{Q}}_{m - 1}}\left( {{\bf{z}}_M(1:m-1) - {\bf{\bar z}}_m } \right) +{{s}_{{p_m}}}{\bf{\tilde w}}_{m-1} -  \\
		\left[ \setlength{\arraycolsep}{2.5pt}
		\begin{array}{*{20}{c}}
			{\bf{Q}}_{m - 1}^{} + \omega _{m}^{}
			{\bf{\tilde w}}_{m-1}^{} {\bf{\tilde w}}_{m-1}^H & - \omega _{m} {\bf{\tilde w}}_{m-1}
		\end{array} \right] \left[ {\setlength{\arraycolsep}{0.1pt}
			\begin{array}{*{20}{c}}
				{{\bf{z}}_M(1:m-1) - {\bf{\bar z}}_m}\\
				{ {\bf{z}}_M(m) - {{\bf{z}}_m}(m)}
		\end{array}} \right].
	\end{multline*}
\end{small}%

After simplification, the above expression becomes

			\begin{multline}\label{dd2wwwzz2390kds32}
				{{\bf{d}}_{m - 1}} - {\bf{\bar d}}_m  = {{s}_{{p_m}}}{\bf{\tilde w}}_{m-1} +
				{\bf{\tilde w}}_{m-1} \\
				\times\left[ {\begin{array}{*{20}{c}}
						- \omega _{m} {\bf{\tilde w}}_{m-1}^H  & {\omega _m}
				\end{array}} \right] \left[ {\begin{array}{*{20}{c}}
						{{\bf{z}}_M(1:m-1) - {\bf{\bar z}}_m}\\
						{  {\bf{z}}_M(m) - {{\bf{z}}_m}(m)}
				\end{array}} \right],
			\end{multline}

because all terms involving ${{\bf{Q}}_{m - 1}}$ cancel each other. Finally, substituting (\ref{d2wpsizz239ds32d}) and ${\bf{\tilde w}}_{m-1} = -{\bf{w}}_{m-1}/\omega_{m}$ (following from (\ref{derive3a})) into (\ref{dd2wwwzz2390kds32}) yields
${{\bf{d}}_{m - 1}} - {{\bf{\bar d}}_m} = - {s_{{p_m}}}\frac{{{{{\bf{w}}}_m}}}{{{\omega_m}}} - \frac{{{{{\bf{w}}}_m}}}{{{\omega_m}}}{{\bf{d}}_m}(m)$,
which is equivalent to (\ref{dUpdate2388932sde3}). In the above derivation, ${\bf{z}}_M(1:m) - {{\bf{z}}_m}$ in (\ref{dminus1dse32sd32before}) and (\ref{d2wpsizz239ds32d}) is partitioned as
$\left[ {\begin{array}{*{20}{c}}
{{\bf{z}}_M(1:m-1) - {\bf{\bar z}}_m}\\
{{\bf{z}}_M(m) - {{\bf{z}}_m}(m)}
\end{array}} \right]$.
As an alternative to ${{\bf{d}}_m}$ defined in (\ref{defd329023dks32d3}), introduce

\begin{equation}\label{t2Qz23sd23ds}
{{\mathbf{t}}_{m}}={{\mathbf{Q}}_{m}}{{\mathbf{z}}_{m}}.
\end{equation}

Substituting (\ref{t2Qz23sd23ds}) into (\ref{defd329023dks32d3}) establishes the following relationship between ${{\bf{d}}_m}$ and ${{\mathbf{t}}_{m}}$:

\begin{equation}\label{d2Qzt934dscerwsdf34}
{{\mathbf{d}}_{m}}={{\mathbf{Q}}_{m}}{{\mathbf{z}}_{M}}(1:m)-{{\mathbf{t}}_{m}}.
\end{equation}

Accordingly, (\ref{equ:16Update}) and (\ref{dUpdate2388932sde3}) can be equivalently replaced by

\begin{equation}\label{s2t239032sd23s}
{{\hat s}_{{p_m}}} = {{\bf{t}}_m}(m)
\end{equation}

and

\begin{equation}\label{t2tstq349sd23ds23}
{{\bf{t}}_{m - 1}} = {{\bf{\bar t}}_m} + \left( {{s_{{p_m}}} - {{\bf{t}}_m}(m)} \right){{\bf{w}}_m}/{\omega _m},
\end{equation}

respectively, where ${{\bf{\bar t}}_m}$ is obtained by removing the last entry of ${{\bf{t}}_m}$. The derivations of (\ref{s2t239032sd23s}) and (\ref{t2tstq349sd23ds23}) are provided in Appendix B. The proposed recursive algorithm using ${{\bf{t}}_m}$ has been extended to the recursive ISIC detector in \cite{ZhuRethinkingISIC}. Nevertheless, this formulation is not considered in the remainder of this paper because it requires slightly more computations\footnote{The initialization of ${{\mathbf{t}}_{M}}$ by (\ref{t2Qz23sd23ds}) requires $M^{2}$ multiplications and additions, whereas that of ${{\bf{d}}_{M}}$ by (\ref{d20wesddswe3}) does not require any calculations. On the other hand, compared with (\ref{equ:16Update}), (\ref{s2t239032sd23s}) saves only $\sum\limits_{m = 1}^M m \approx \frac{{{M^2}}}{2}$ multiplications and additions in total.} when (\ref{s2t239032sd23s}) and (\ref{t2tstq349sd23ds23}) are used in place of (\ref{equ:16Update}) and (\ref{dUpdate2388932sde3}), respectively.

\subsection{PROPOSED RECURSIVE ALGORITHM WITH BOTH SPEED ADVANTAGE AND MEMORY SAVING}
In the proposed recursive algorithm with speed advantage, i.e., \textbf{Algorithm 1} with (\ref{derive2Globecomm2005all3}) replaced by (\ref{derive3}), (\ref{equ:16}) and (\ref{equ:BLAST_IC}) from Improvement \uppercase\expandafter{\romannumeral+2} can be replaced by (\ref{equ:16Update}) and (\ref{dUpdate2388932sde3}) from Improvement \uppercase\expandafter{\romannumeral+6}, respectively. Consequently, ${{\bf{v}}_{m-1}}$ used in (\ref{equ:BLAST_IC}) is replaced by ${\bf{w}}_{m-1}$ and $\omega_{m}$ in (\ref{dUpdate2388932sde3}), which constitute ${\bf{q}}_m$ in (\ref{equ:16Update}). This replacement eliminates the need for ${{\bf{R}}_{M}}$ during the \textsl{recursion} phase. Therefore, in the \textsl{initialization} phase, the memory region occupied by ${{\bf{R}}_{M}}$ can subsequently be reused to store ${{\bf{Q}}_{M}}$, thereby reducing the memory requirement.

When ${{\bf{Q}}_{M}}$ is computed from ${{\bf{R}}_{M}}$ through the iterative application of (\ref{equQsmally26}) and (\ref{derive3}), ${\bf{Q}}_i$ is obtained from ${\bf{Q}}_{i-1}$ and $i$-th column of ${\bf{R}}_i$, i.e., ${\bf{v}}_{i}$ and $\gamma_{i}$, at the $i$-th iteration for $2 \leq i \leq M$. After this iteration, only columns $i+1$ to $M$ in the upper triangular part of ${\bf{R}}_M$ are required for the subsequent iterations. Therefore, overwriting the submatrix ${\bf{R}}_i$ in ${\bf{R}}_M$ with ${\bf{Q}}_i$ does not affect the remaining computations. For this in-place implementation, (\ref{derive3}) from Improvement \uppercase\expandafter{\romannumeral+5} is rewritten as

\begin{subnumcases}{\label{EquR2QsaveMem83s3d}}
{\bf{\tilde q}} = {\bf{R}}(1:i - 1,1:i - 1){\bf{R}}(1:i - 1,i), &  \label{EquR2QsaveMem83s3d1}\\
{\bf{R}}(i,i) = 1/\left( {{\bf{R}}(i,i) - {\bf{R}}{{(1:i - 1,i)}^H}{\bf{\tilde q}}} \right), &  \label{EquR2QsaveMem83s3d2}\\
{\bf{R}}(1:i - 1,i) =  - {\bf{R}}(i,i){\bf{\tilde q}},  &  \label{EquR2QsaveMem83s3d3}\\
{\bf{R}}(i,1:i - 1) = {\bf{R}}{(1:i - 1,i)^H},   &  \label{EquR2QsaveMem83s3d4}\\
\begin{array}{l}
{\bf{R}}(1:i - 1,1:i - 1) =   \\
{\bf{R}}(1:i - 1,1:i - 1) - {\bf{\tilde q}}{\bf{R}}(i,1:i - 1).
\end{array} & \label{EquR2QsaveMem83s3d5}
\end{subnumcases}

Specifically, (\ref{get_u_less_div}), (\ref{derive3b}), and (\ref{derive3a}) are expressed as (\ref{EquR2QsaveMem83s3d1}), (\ref{EquR2QsaveMem83s3d2}), and (\ref{EquR2QsaveMem83s3d3}), respectively. In (\ref{EquR2QsaveMem83s3d4}), the conjugate transpose of column $i$ in the upper triangular part of ${\bf{Q}}_i$ is used to overwrite row $i$ in its lower triangular part. Moreover, substituting (\ref{derive3a}) into (\ref{derive3c}) gives ${\bf{\bar Q}}_{m-1}^{} = {\bf{Q}}_{m-1}^{} - {\bf{\tilde w}}_{m-1} {\bf{w}}_{m-1}^H$, which is implemented in (\ref{EquR2QsaveMem83s3d5}).

To overwrite ${\bf{R}}_M$ with ${\bf{Q}}_M$, (\ref{EquR2QsaveMem83s3d}) is applied iteratively for $i=2,3,\cdots,M$. Before these iterations, ${\bf{R}}_1$ is overwritten by ${\bf{Q}}_1$ according to

\begin{equation}\label{H112H11de238jds}
{\bf{R}}(1,1) = 1/{\bf{R}}(1,1),
\end{equation}

which follows directly from (\ref{Q1compute298jd32}). It is worth noting that, in (\ref{EquR2QsaveMem83s3d}), only column $i$ in the upper triangular part of ${\bf{R}}_i$ is required to compute ${\bf{Q}}_i$, whereas the entire ${\bf{R}}_i$ is overwritten by ${\bf{Q}}_i$.





During the \textsl{initialization} phase, the memory occupied by ${{\bf{H}}_{M}}$ is reused to store ${{\bf{R}}_{M}}$, since ${{\bf{H}}_{M}}$ is no longer required in the \textsl{recursion} phase. Specifically, we employ an implementation that overwrites the upper triangular part of a square submatrix of

\begin{equation}\label{HtransStore30ds32}
{\bf{\tilde H}}={\bf{H}}_{M}^H
\end{equation}

with the corresponding upper triangular part of ${{\bf{R}}_{M}}$. Substituting (\ref{HtransStore30ds32}) into (\ref{Rm2Hm3298kds32}) gives
${\bf{R}}_M = {\bf{\tilde H}}{\bf{\tilde H}}^H + \alpha {\bf{I}}_{M}$.
Accordingly, row $i$ in the upper triangular part of the square submatrix ${\bf{\tilde H}}(:,1:M)$ is overwritten by the corresponding entries of ${\bf{R}}_M$ according to

\begin{equation}\label{H2RAug9aewsd}
{\bf{\tilde H}}(i,i:M) = {\bf{\tilde H}}(i,:){\bf{\tilde H}}{(i:M,:)^H}+\left[\alpha \enspace {{\bf{0}}_{M - i}^T}\right].
\end{equation}

By evaluating (\ref{H2RAug9aewsd}) successively for $i=1, 2, \cdots, M$, the upper triangular part of the square submatrix ${\bf{\tilde H}}(:,1:M)$ is fully overwritten by that of ${\bf{R}}_M$. This in-place memory reuse does not affect the subsequent computations because only rows $i$ to $M$ of ${\bf{\tilde H}}$ are required to calculate row $i$ of the upper triangular part of ${\bf{R}}_M$. Hence, after the $i$-th iteration, row $i$ of ${\bf{\tilde H}}$ is no longer needed in the $(i+1)$-th through $M$-th iterations of (\ref{H2RAug9aewsd}). Moreover, the initial ${{\bf{d}}_{M}}$ is set according to (\ref{d20wesddswe3}). Substituting (\ref{HtransStore30ds32}) into (\ref{zm2HmxmDef3290sd23}) with $m=M$ allows ${\bf{z}}_M$ to be computed directly from ${\bf{\tilde H}}$ as

\begin{equation}\label{z2Hwave9230dsk32}
{\bf{z}}_M  = {\bf{\tilde H}} {\bf{x}}^{(M)}.
\end{equation}


During the \textsl{recursion} phase, ${\bf{Q}}_{m}$ is still permuted according to the SNR-based ordering. Consequently, ${\bf{q}}_m$, consisting of ${\bf{w}}_{m-1}$ and $\omega_{m}$, is permuted accordingly. It follows from (\ref{equ:16Update}) and (\ref{dUpdate2388932sde3}) that entries $l_m$ and $m$ of ${\bf{z}}_M$ and ${{\bf{d}}_{m}}$ must also be interchanged consistently.


The proposed recursive V-BLAST algorithm combining both speed advantage and memory saving is summarized in \textbf{Algorithm 3}. It is developed from \textbf{Algorithm 1}, i.e., the improved algorithm with speed advantage in \cite{zhf6}, by replacing Improvements \uppercase\expandafter{\romannumeral+1} and \uppercase\expandafter{\romannumeral+2} with Improvements \uppercase\expandafter{\romannumeral+5} and \uppercase\expandafter{\romannumeral+6}, respectively. Specifically, \textbf{Algorithm 3} employs (\ref{EquR2QsaveMem83s3d}) from Improvement \uppercase\expandafter{\romannumeral+5}, (\ref{equ:16Update}) and (\ref{dUpdate2388932sde3}) from Improvement \uppercase\expandafter{\romannumeral+6}, together with (\ref{d20wesddswe3}) and (\ref{H112H11de238jds})--(\ref{z2Hwave9230dsk32}).

\begin{algorithm}[!t]
\caption{Proposed Recursive V-BLAST Algorithm with Both Speed Advantage and Memory Saving}
\renewcommand{\algorithmicrequire}{\textbf{Initialization:}} 
\renewcommand{\algorithmicensure}{\textbf{Recursion:}}
\label{alg:Framwork}
\begin{algorithmic}[1] %
\Require Set
${\bf{p}} = \left[ 1, 2, \cdots,
M \right]^T$ and ${{\bf{d}}_{M}}={{\bf{0}}_{M}}$; 
Store ${\bf{\tilde H}}={\bf{H}}^H$
and compute
$ {\bf{z}}_M  = {\bf{\tilde H}} {\bf{x}}$;  
Compute
(\ref{H2RAug9aewsd}) iteratively for $i=1, 2, \cdots, M$ to cover the upper triangular part of ${\bf{\tilde H}}(:,1:M)$
with that of ${\bf{R}}_M$;
Compute  
(\ref{H112H11de238jds}), and then compute
(\ref{EquR2QsaveMem83s3d})
iteratively for $i=2,3,\cdots,M$, to cover
the entire matrix    ${\bf{R}}_M$ (i.e., the entire square submatrix ${\bf{\tilde H}}(:,1:M)$)
with
${\bf{Q}}_M$;
\Ensure For $m=M,M-1,\cdots,2$:
\State Find $ l_m  = \mathop {\arg \min }\limits_{j = 1,2...}^m {q}_{jj} $, where
$q_{jj}$
is the $j$-th diagonal entry of  ${\bf{Q}}_{m}$;
Permute entries $l_m$ and $m$ in ${\bf{p}}$, ${\bf{z}}_M$  and  ${{\bf{d}}_{m}}$;
Permute rows and columns $l_m$
and $m$ in ${\bf{Q}}_{m}$;
\State Compute $\hat s_{{p_m}}$ by (\ref{equ:16Update}), which is quantized
to  $s_{{p_m}}$;
\State  Deflate ${\bf{d}}_m$  to ${{\bf{d}}_{m - 1}} $ by
(\ref{dUpdate2388932sde3}), and deflate
${{\bf{Q}}_{m}}$ to ${{\bf{Q}}_{m-1}}$ by (\ref{equ:14});
\renewcommand{\algorithmicrequire}{\textbf{Solution:}} 
\Require When $m = 1$, only run step 2 to get $s_{p_1}$.
\end{algorithmic}
\end{algorithm}

\section{PROPOSED HYBRID RECURSIVE AND SQUARE-ROOT V-BLAST DETECTION}

To the best of our knowledge, the inverse-Cholesky-factor-based square-root algorithm in \cite{refZHFtrans2011} has the lowest computational cost among the existing square-root algorithms\footnote{The square-root algorithm in \cite{refZHFtrans2015} and that in \cite{refZHFtrans2011} require the same complexity, as mentioned in \cite{zhfcompare2022sqrt}.}, and it also requires fewer computations than the recursive algorithm with speed advantage in \cite{zhf6}, i.e., \textbf{Algorithm 1}. Compared with the two proposed recursive algorithms, however, the square-root algorithm in \cite{refZHFtrans2011} has lower best-case and average complexity but higher worst-case complexity\footnote{The relevant complexity comparison is presented in the next section. The complexity of an algorithm is commonly characterized in terms of its best-case, average-case, and worst-case complexity.}, because both proposed recursive algorithms are computationally more efficient than the recursive algorithm with speed advantage in \cite{zhf6}. This complementary complexity behavior motivates us to further improve the proposed recursive algorithm with both speed advantage and memory saving by switching to the inverse-Cholesky-factor-based square-root algorithm under appropriate conditions. Accordingly, a hybrid scheme combining recursive and square-root steps is proposed in this section.

\subsection{INVERSE-CHOLESKY SQUARE-ROOT V-BLAST AND ITS MEMORY-SAVING VARIANT}

The square-root algorithm in \cite{refZHFtrans2011} operates on the inverse Cholesky factor of ${\bf{R}}_m$ defined in (\ref{Rm2Hm3298kds32}). Specifically, it employs the upper-triangular matrix ${\bf{F}}_{m}$ satisfying

\begin{equation}{\label{ToBeSubstedforGamaDefFm}}
{{\bf{F }}_{m}}  {{\bf{F }}_{m} ^{H}}   = {\bf{R}}_{m}^{ - 1}.   
\end{equation}

During the initialization phase, ${\bf{F}}_{m}$ is recursively constructed from ${\bf{F}}_{m-1}$ for $m=1,2,\cdots,M$ according to

\begin{subnumcases}{\label{equ20_19}}
\lambda _m  = 1/\sqrt {{{\gamma _m  - {\bf{v}}_{m - 1}^H
{\bf{F }}_{m-1}{{\bf{F }}_{m-1}^H}
{\bf{v}}_{m - 1} }}}, &\label{equ:20}\\
{\bf{u}}_{m - 1}  =  - \lambda _m {\bf{F }}_{m-1} {{\bf{F
}}_{m-1}^H}{\bf{v}}_{m - 1}, &\label{equ:19}\\
{\bf{F }}_{m}  = \left[ {\begin{array}{*{20}c}
		{{\bf{F }}_{m-1} } & {{\bf{u}}_{m - 1} }  \\
		{{\bf{0}}_{m - 1}^T } & {\lambda _m }  \\
\end{array}} \right], &\label{equ:13}
\end{subnumcases}

where $\gamma_m$ and ${\bf{v}}_{m - 1}$ are the corresponding components of ${\bf{R}}_{m}$ specified in (\ref{equ:11}). The recursion is initialized by

\begin{equation}\label{equ:22}
{\bf{F }}_{1}  = \sqrt {{{\bf{R}}_{1}^{ - 1} }},
\end{equation}

and continues until the initial ${\bf{F}}_{M}$ is obtained.

The square-root algorithm in \cite{refZHFtrans2011} also contains an iterative detection phase. At each iteration, the row with the minimum length in the upper-triangular ${\bf{F}}_{m}$ ($m=M,M-1,\dots,2$) is permuted to the last row. This row can be represented as the $i$-th ($i \leq m$) row, whose first $(i-1)$ entries are zeros. The permuted ${\bf{F}}_{m}$ is subsequently block upper-triangularized through a unitary transformation ${\bf{\Sigma}}$, i.e.,

\begin{equation}\label{equ:9}
{\bf{F }}_{m} {\bf{\Sigma }} = \left[ {\begin{array}{*{20}c}
{{\bf{F }}_{m-1} } & {{\bf{u}}_{m - 1} }  \\
{{\bf{0}}_{m - 1}^T } & {\lambda _m }  \\
\end{array}} \right].
\end{equation}

The quantities ${{\bf{u}}_{m - 1} }$ and ${\lambda_m}$ in (\ref{equ:9}) are then used to construct the linear MMSE estimate

\begin{equation} \label{equ:24}
\hat s_{p_m }   = \lambda _m   \left[ {\begin{array}{*{20}c}
{{{\bf{u}}_{m - 1}^H }  } & {{\lambda _m }  }  \\
\end{array}} \right]  {\bf{z}}_{m}.
\end{equation}

The transformation ${\bf{\Sigma}}$ in (\ref{equ:9}) can be implemented using only $(m-i)$ Givens rotations \cite{zhf_VTC08_6}, namely,

\begin{equation}\label{equ:26}
	{\bm \Sigma}^g_m = {\bf{\Omega }}_{i,i + 1}^i {\bf{\Omega }}_{i +
		1,i + 2}^i \cdots {\bf{\Omega }}_{m - 1,m}^i=\prod\limits_{j = i}^{m
		- 1} {{\bf{\Omega }}_{j,j + 1}^i }.
\end{equation}

In (\ref{equ:26}), ${\bf{\Omega}}_{k,n}^i$ rotates the $k$-th and $n$-th entries of each row of ${\bf{F}}_{m}$ and eliminates the $k$-th entry of the $i$-th row. Therefore, the computational cost of (\ref{equ:26}) is low when $m-i$ is small and becomes zero in the best case where $m-i=0$. In the subsequent iteration with $(m-1)$ undetected symbols, ${{\bf{F}}_{m-1} }$ in (\ref{equ:9}) is retained, while ${\bf{z}}_{m-1}$ is obtained by applying (\ref{equ:BLAST_IC}) to cancel the interference of $s_{p_m}$ in the permuted ${\bf{z}}_m$.


Since (\ref{equ:BLAST_IC}) requires ${{\bf{v}}_{m-1}}$ from ${\bf{R}}_m$, we develop a memory-saving variant of the square-root algorithm in \cite{refZHFtrans2011} that eliminates the use of ${\bf{R}}_M$ during the iterative detection phase. First, substituting (\ref{equ:4}) into (\ref{ToBeSubstedforGamaDefFm}) gives

\begin{equation}\label{ApdxJ21equ1}
{\bf{F}}_m{\bf{F}}_m^H={\bf{Q}}_{m}.
\end{equation}

Further substituting (\ref{equQsmally26}) and (\ref{equ:9}) into (\ref{ApdxJ21equ1}) yields

\begin{subnumcases}{\label{omega2lamda111}}
{\omega _{m} }= {\lambda _m^2 },	  & \label{omega2lamda111aa}\\
{{\bf{w}}_{m-1} }= \lambda _m {{\bf{u}}_{m - 1} },  & \label{omega2lamda111bb} \\
{\bf{q}}_m^H  = \lambda _m   \left[ {\begin{array}{*{20}c}
{{{\bf{u}}_{m - 1}^H }  } & {{\lambda _m }  }  \\
\end{array}} \right].  &  \label{equ:24To2qm}
\end{subnumcases}

Substituting (\ref{omega2lamda111aa}) and (\ref{omega2lamda111bb}) into (\ref{dUpdate2388932sde3}) then gives

\begin{equation}\label{dUpdate2388932sde3NewY26}
{{\bf{d}}_{m - 1}} = {\bf{\bar d}}_m  - \left( {{{s}_{{p_m}}}+{\bf{d}}_m^{}(m)} \right) {{\bf{u}}_{m - 1} }/{\lambda _m },
\end{equation}

whereas substituting (\ref{equ:24To2qm}) into (\ref{equ:16Update}) yields

\begin{equation}\label{equ:16UpdateNewY26}
\hat s_{p_m }  = \lambda _m   \left[ {\begin{array}{*{20}c}
{{{\bf{u}}_{m - 1}^H }  } & {{\lambda _m }  }  \\
\end{array}} \right]  {\bf{z}}_M(1:m) - {{\bf{d}}_m}(m).
\end{equation}

Accordingly, a memory-saving version of the square-root V-BLAST algorithm in \cite{refZHFtrans2011} is obtained by replacing the computation of $\hat s_{p_m}$ using (\ref{equ:24}) with that using (\ref{equ:16UpdateNewY26}), and by replacing the update of ${\bf{z}}_m$ through (\ref{equ:BLAST_IC}) with the update of ${\bf{d}}_m$ through (\ref{dUpdate2388932sde3NewY26}).





\subsection{HYBRID SWITCHING BETWEEN RECURSIVE AND SQUARE-ROOT UPDATES}

In this subsection, we propose a hybrid scheme that adaptively switches between the proposed recursive algorithm with both speed advantage and memory saving and the proposed memory-saving square-root algorithm. The resulting scheme substantially reduces the best-case complexity of the recursive algorithm and the worst-case complexity of the square-root algorithm, while also providing a slight reduction in average complexity compared with both algorithms.


To enable switching between the recursive and square-root steps, the proposed hybrid scheme propagates a $m \times {M'}$ matrix ${\bf{F}}_{m \times {M'}}$ ($m \leq {M'} \leq M$) that contains some or all rows of the inverse Cholesky factor ${\bf{F}}_{{M'}}$ satisfying (\ref{ApdxJ21equ1}). The initial inverse Cholesky factor ${\bf{F}}_{m \times {M'}} = {\bf{F}}_{{M}}$ ($m = {M'} = M$) can be obtained by computing (\ref{equ:22}) and then applying (\ref{equ20_19}) iteratively for $m=2,3,\cdots,M$. Since the diagonal entries of ${\bf{Q}}_{M}$ are not directly available from ${\bf{F}}_M$, they must be separately maintained for determining the optimal detection order. To this end, let ${\bf{e}}_{M}$ denote the column vector containing all $M$ diagonal entries of ${\bf{Q}}_{M}$. Its $i$-th ($i=1,2,\cdots,M$) entry is calculated as

\begin{equation}\label{eMcomputeJul8}
{\bf{e}}_{M}(i) = {\bf{F}}_M(i,i:M) {\bf{F}}_M^H(i,i:M).
\end{equation} 

At recursion $m$ ($m=M,M-1,\cdots,2$), ${\bf{e}}_{m}$ is updated according to

\begin{equation}\label{ApdxJ22equ4recurJul8}
{\bf{e}}_{m-1} = {\bf{\bar e}}_{m}   - \frac{1}{\omega _{m}} \left| {{\bf{w}}_{m-1}^{}} \right| \odot \left| {{\bf{w}}_{m-1}^{}} \right|
\end{equation}

when the recursive step is selected, whereas

\begin{equation}\label{ApdxJ22equ4sqrtEupdate}
{\bf{e}}_{m-1} = {\bf{\bar e}}_{m}   -  \left|  {{\bf{u}}_{m - 1} }  \right| \odot \left|  {{\bf{u}}_{m - 1} }  \right|
\end{equation}

is used when the square-root step is selected. Equations (\ref{ApdxJ22equ4recurJul8}) and (\ref{ApdxJ22equ4sqrtEupdate}) follow from (\ref{equ:14}) and (\ref{equ:9}), respectively. Here, $\odot$ denotes the Hadamard product, and ${\bf{\bar e}}_{m}$ is obtained by removing the last entry from the permuted ${\bf{e}}_{m}$.


Initialize $m=M$, ${M'}=M$, and ${\bf{F}}_{m \times {M'}}={\bf{F}}_M$. The proposed hybrid scheme is then described through two cases: one involving the square matrix ${\bf{F}}_{m \times {M'}}={\bf{F}}_m$ with $M'=m$, and the other involving the non-square matrix ${\bf{F}}_{m \times {M'}}$ with $M' \neq m$, where $M' \geq m \geq 2$. At recursion $m$ ($M \geq m \geq 2$), define $ l_m \triangleq \mathop {\arg \min }\limits_{j = 1,2...}^m {\bf{e}}_{m}(j)$. To determine whether the recursive or square-root step should be performed, simple switching criteria are adopted. Specifically, $l_m/m$ is used when $M'=m$, whereas $m/M'$ is used when $M' \neq m$. These switching criteria are detailed in the remainder of this subsection.







We first consider the case where $M'=m$. When $l_m/m$ is relatively large, only a small number (i.e., $m-l_m$) of Givens rotations are required in (\ref{equ:26}), resulting in a relatively low computational cost. In this case, the proposed hybrid scheme selects the proposed memory-saving square-root algorithm. Specifically, the block upper-triangularized ${\bf{F}}_{m}$, ${{\bf{d}}_{m - 1}}$, and $\hat s_{p_m}$ are computed using (\ref{equ:9}), (\ref{dUpdate2388932sde3NewY26}), and (\ref{equ:16UpdateNewY26}), respectively. Meanwhile, ${\bf{F}}_{m-1}={\bf{F}}_{(m-1) \times {M'}}$ (with ${M'}=m-1$) and ${\bf{e}}_{m-1}$ required for the next recursion are obtained from (\ref{equ:9}) and (\ref{ApdxJ22equ4sqrtEupdate}), respectively. In contrast, when $l_m/m$ is relatively small, the proposed hybrid scheme selects the recursive step, as described below.

To derive the recursive step, we first verify in Appendix C that ${\bf{Q}}_m$ (obtained by recursively deflating ${\bf{Q}}_{M'}$ through (\ref{equ:14})) satisfies

\begin{equation}\label{ApdxJ21equ7}
{\bf{Q}}_{m}  = {\bf{F}}_{m \times M'} {\bf{F}}_{m \times M'}^H  - {\bf{W}}_{M'-m} {\bf{C}}_{M'-m} {\bf{W}}_{M'-m}^H,
\end{equation}

where the $m \times (M'-m)$ matrix ${\bf{W}}_{M'-m}$ is defined as

\begin{equation}\label{ApdxJ21equ5}
{\bf{W}}_{M'-m}  = \left[ {\begin{array}{*{20}c}
{{\bf{w}}_{M' - 1}^{\bcancel{M' - 1-m}} } & 	{{\bf{w}}_{M' - 2}^{\bcancel{M' - 2-m}} } &  \cdots   & {{\bf{w}}_{m}^{} }  \\
\end{array}} \right],
\end{equation}

and the diagonal matrix ${\bf{C}}_{M'-m}$ is defined as

\begin{equation}\label{ApdxJ21equ6}
{\bf{C}}_{M'-m}  = \mathrm{diag}(1/\omega_{M'},1/\omega_{M' - 1}, \cdots,  1/\omega_{m+1}).
\end{equation}

In (\ref{ApdxJ21equ5}), ${{\bf{w}}_{i}^{\bcancel{i-m}} }$ ($M'-1 \geq i \geq m$) denotes the permuted ${\bf{w}}_{i}$ after its last $(i-m)$ entries have been removed, and therefore contains $i-(i-m)=m$ entries. It follows from (\ref{ApdxJ21equ5}) and (\ref{ApdxJ21equ6}) that, for $m \leq M'-2$, ${\bf{W}}_{M'-m}$ and ${\bf{C}}_{M'-m}$ can be recursively updated as

\begin{equation}\label{ApdxJ21equ5UpdateY26}
{\bf{W}}_{M'-m+1}  = \left[ {\begin{array}{*{20}c}
{\bf{\bar W}}_{M'-m}    & {{\bf{w}}_{m-1}^{} }  \\
\end{array}} \right]
\end{equation}

and 

\begin{equation}\label{ApdxJ21equ6UpdateY26}
{\bf{C}}_{M'-m+1}  = \left[ {\begin{array}{*{20}c}
{\bf{C}}_{M'-m}   & 0  \\
0  & {1/\omega _{m} }  \\
\end{array}} \right],
\end{equation}

respectively, where ${\bf{\bar W}}_{M'-m}$ is obtained by removing the last row from the permuted ${\bf{W}}_{M'-m}$. If the hybrid scheme switches to the square-root step in a subsequent recursion, some rows of ${\bf{Q}}_{m}$ in (\ref{ApdxJ21equ7}) are never used. To avoid these unnecessary computations, (\ref{ApdxJ21equ7}) is employed only to calculate the $m$-th row of the permuted ${\bf{Q}}_{m}$ required at recursion $m$ ($m=M',M'-1,\cdots,1$), namely,

\begin{align}\label{ApdxJ21equ10bestJul7}
{\bf{q}}_{m}^H &= {\bf{F}}_{m \times M'}(m,:) {\bf{F}}_{m \times {M'}}^H \notag\\
& - {\bf{W}}_{{M'}-m}(m,:) {\bf{C}}_{{M'}-m} {\bf{W}}_{{M'}-m}^H,
\end{align}%

where ${\bf{F}}_{m \times M'}(m,:)$ and ${\bf{W}}_{{M'}-m}(m,:)$ denote the $m$-th (i.e., last) rows of the permuted ${\bf{F}}_{m \times {M'}}$ and ${\bf{W}}_{{M'}-m}$, respectively.

When the recursive step is selected, ${\bf{q}}_{m}^H$ is first evaluated using (\ref{ApdxJ21equ10bestJul7}) and then substituted into (\ref{equ:16Update}) to obtain $\hat s_{{p_m}}$. For the subsequent recursion, ${\bf{q}}_{m}^H = [{{\bf{w}}_{m-1}^H }, {\omega_{m} }]$, as follows from (\ref{equQsmally26}), is used to update ${\bf{d}}_m$, ${\bf{e}}_{m}$, ${\bf{W}}_{M'-m}$, and ${\bf{C}}_{M'-m}$ to ${{\bf{d}}_{m - 1}}$, ${\bf{e}}_{m-1}$, ${\bf{W}}_{M'-m+1}$, and ${\bf{C}}_{M'-m+1}$ through (\ref{dUpdate2388932sde3}), (\ref{ApdxJ22equ4recurJul8}), (\ref{ApdxJ21equ5UpdateY26}), and (\ref{ApdxJ21equ6UpdateY26}), respectively. In addition, ${\bf{F}}_{m \times M'}$ is reduced to ${\bf{F}}_{(m-1) \times M'}$ by deleting the last row of the permuted ${\bf{F}}_{m \times M'}$.


%
%

After a recursive step, ${\bf{F}}_{m \times M'}$ becomes a non-square matrix with ${M'} \neq m$. The remainder of this subsection therefore focuses on this case. Suppose that, at recursion $m$ with the initial ${M'} \neq m$, a square matrix ${\bf{F}}_m={\bf{F}}_{m \times M'}$ (with $M'=m$) is computed for the remaining $m$ undetected antennas. Then, both ${\bf{W}}_{M'-m}$ and ${\bf{C}}_{M'-m}$ are reset to empty matrices. As a result, the computational cost of evaluating (\ref{ApdxJ21equ10bestJul7}) in the subsequent recursive steps is reduced, and further complexity reduction becomes possible by switching to the square-root step. Accordingly, at each recursion initialized with ${M'} \neq m$, the proposed hybrid scheme compares the total complexity required to compute ${\bf{F}}_m$ with the complexity reduction achievable by using ${\bf{F}}_m$ in the subsequent steps. The square matrix ${\bf{F}}_m$ is computed whenever the former does not exceed the latter.


%
%

\begin{algorithm}[!t]
\caption{Proposed Hybrid V-BLAST Algorithm Switching Between the Recursive and Square-Root Steps}
\renewcommand{\algorithmicrequire}{\textbf{Initialization:}} 
\renewcommand{\algorithmicensure}{\textbf{Recursion:}}
\label{alg:Framwork}
\begin{algorithmic}[1] %
\Require Set
${\bf{p}} = \left[ 1, 2, \cdots,
M \right]^T$ and ${{\bf{d}}_{M}}={{\bf{0}}_{M}}$; 
Compute
${\bf{\tilde z}}_M   = {\bf{H}}^H {\bf{x}}$ and  ${\bf{R}}_M = {\bf{H}}^H {\bf{H}} + \alpha {\bf{I}}_{M} $;  
Obtain ${\bf{F}}_M$ by computing (\ref{equ:22}) and then computing
(\ref{equ20_19}) iteratively; 
Obtain 	${\bf{e }}_M$ by computing (\ref{eMcomputeJul8})
 for $1 \le i \le M$;
 Initialize  ${M'} = M$, $m = M$ and ${\bf{F}}_{m \times  {M'}}={\bf{F}}_M$;  Set
${\bf{W}}_{M'-m}$ and ${\bf{C}}_{M'-m}$ to empty matrices;
\Ensure For $m=M,M-1,\cdots,2$:
\State    	Find $ l_m  = \mathop {\arg \min }\limits_{i = 1,2\cdots}^m {\bf{e }}_m (i) $; Let
$ {p_m} ={\bf{p}}(l_m ) $,  and  delete ${\bf{p}}(l_m ) $ in ${\bf{p}}$;
Permute ${\bf{e }}_m$,	${\bf{\tilde z}}_m$   and ${{\bf{d}}_{m}}$ by deleting the $l_m$-th entries and  
moving them to the end; Permute ${\bf{F}}_{m \times  {M'}}$  by deleting the $l_m$-th row and moving it to the bottom;
\If{\emph{$m={M'}$}}
\If{\emph{$l_m / m > \tau $ ($e.g., \tau = 1/2 $)}}
\State  If $l_m \neq m$, block upper-triangularize 
${\bf{F}}_{m \times  {M'}} = {\bf{F}}_m$ by (\ref{equ:9}) with 
${\bf{\Sigma }} $ to be ${\bm \Sigma}^g_m$ in (\ref{equ:26});
\State   Set $Switch2SqrtBestCase = 1$;
\Else[\emph{$l_m /m \le \tau $}]
\State  Set $Switch2SqrtBestCase = 0$;   
\EndIf
\Else[\emph{$m \neq {M'}$    }]
\If{\emph{$m/M' \le \beta $ ($e.g., \beta = 3/5 $) }} 
\State  Set ${M'}=m$, and set
${\bf{W}}_{M'-m}$ and ${\bf{C}}_{M'-m}$ to empty matrices; Obtain  ${\bf{F}}_{m \times  {M'}} = {\bf{F}}_m$ corresponding to  ${\bf{R}}([{\bf{p}} \;  p_m],[{\bf{p}} \; p_m])$
by using the initial ${\bf{F}}_1$ in  (\ref{equ:22}) to 
compute 
(\ref{equ20_19}) iteratively;
\State  Set  $Switch2SqrtBestCase = 1$;
\Else[\emph{$m/M' > \beta $}] 
\State   Permute ${\bf{W}}_{M'-m}$  by deleting the $l_m$-th row and moving it to the bottom;
\State  Set $Switch2SqrtBestCase = 0$;
\EndIf
\EndIf
\If{\emph{$Switch2SqrtBestCase = 1$}}
\State   Compute  $\hat s_{p_m }$ by (\ref{equ:16UpdateNewY26}) and  then quantize $\hat s_{p_m }$
to  $s_{{p_m}}$; Compute ${{\bf{d}}_{m - 1}}$  and ${\bf{e}}_{m-1}$   by  
(\ref{dUpdate2388932sde3NewY26})  and (\ref{ApdxJ22equ4sqrtEupdate}), respectively;
Set $M'=m-1$ and obtain   	
	${\bf{F}}_{(m-1) \times  {M'}} = {\bf{F}}_{m-1}$ 
 in 	(\ref{equ:9});
\Else[\emph{$Switch2SqrtBestCase = 0$}]
\State   Compute  ${\bf{q}}_{m}^H $ and  $\hat s_{{p_m}}$   by   (\ref{ApdxJ21equ10bestJul7}) and  (\ref{equ:16Update}),  respectively, and then quantize $\hat s_{p_m }$ to  $s_{{p_m}}$;  Obtain		
${{\bf{d}}_{m - 1}} $, ${\bf{e}}_{m-1}$, ${\bf{W}}_{M'-m+1}$ and ${\bf{C}}_{M'-m+1}$ by   	(\ref{dUpdate2388932sde3}), (\ref{ApdxJ22equ4recurJul8}),
(\ref{ApdxJ21equ5UpdateY26}) and  (\ref{ApdxJ21equ6UpdateY26}), respectively;
Delete  the last row of ${\bf{F}}_{m \times M'}$ to obtain  ${\bf{F}}_{(m-1) \times M'}$;
\EndIf			
\State   Delete the last entry of   ${\bf{\tilde z}}_m$ to obtain ${\bf{\tilde z}}_{m-1}$;
\renewcommand{\algorithmicrequire}{\textbf{Solution:}} 
\Require When $m = 1$, use (\ref{equ:16Update}) with ${\bf{q}}_1={\bf{e}}_{1}$ to compute $\hat s_{{p_1}}$, and then  quantize  $\hat s_{{p_1}}$
to  $s_{{p_1}}$.
\end{algorithmic}
\end{algorithm}

The proposed hybrid scheme is summarized in \textbf{Algorithm 4}. In the next section, we will discuss the settings of $\tau$ and $\beta$, which are used in the conditions $l_m / m > \tau$ and $m/M' \leq \beta$, respectively. It is worth noting that, when ${\bf{F}}_m$ is computed at a recursion $m$ with the initial $M' \neq m$, the resulting ${\bf{F}}_m$ is the inverse Cholesky factor associated with ${\bf{R}}([{\bf{p}} ; p_m],[{\bf{p}} ; p_m])$\footnote{The notation here follows the MATLAB standard.}, where $p_m$ denotes the index of the antenna detected at recursion $m$. Once ${\bf{F}}_m$ has been obtained, the proposed scheme sets $Switch2SqrtBestCase = 1$ and switches to the square-root step under the best-case condition $m-i=0$ with $i=l_m$, where no Givens rotation is required in (\ref{equ:26}).

\section{COMPARISON OF COMPLEXITY AND MEMORY REQUIREMENT}

\begin{table}[!t]
\centering
\caption{Dominant Complexities of Presented V-BLAST Algorithms}
\begin{tabular}{|c|c|}\hline
Scheme & Dominant Complexity\\\hline
Original Recursive Algorithm \cite{zhf3} & \makecell[c]{$(3M^2N+\frac{2}{3}M^3$, \\ $\frac{5}{2}M^2N+\frac{1}{2}M^3)$}\\\hline
\makecell[c]{Fastest Known \\ Recursive Algorithm Before \cite{zhf6}} & $(\frac{1}{2}M^2N +\frac{4}{3}M^3)$\\\hline
\makecell[c]{Recursive Algorithm \\ with Memory Saving in \cite{zhf6}} & $(2M^2N +\frac{1}{6}M^3)$\\\hline
\makecell[c]{Recursive Algorithm \\ with Speed Advantage in \cite{zhf6}} & $(\frac{1}{2}M^2N+M^3)$\\\hline
\makecell[c]{Square-Root Algorithm \cite{refZHFtrans2011} \\ in Worst Case} & \makecell[c]{$(\frac{1}{2}M^{2}N+\frac{5}{6}M^{3}$, \\ $\frac{1}{2}M^{2}N+\frac{1}{2}M^{3})$}\\\hline
\makecell[c]{Square-Root Algorithm \cite{refZHFtrans2011} \\ in Average Case} & \makecell[c]{$(\frac{1}{2}M^{2}N+\frac{2}{3}M^{3}$, \\ $\frac{1}{2}M^{2}N+\frac{4}{9}M^{3})$}\\\hline
\textbf{Proposed Two Recursive Algorithms} & $(\frac{1}{2}M^2N +\frac{2}{3}M^3)$\\\hline
\makecell[c]{\textbf{Proposed Hybrid Scheme} \\ \textbf{in Worst or Average Case}} & $\leq$$(\frac{1}{2}M^2N+\frac{2}{3}M^3)$\\\hline
\makecell[c]{\textbf{Square-Root Algorithm \cite{refZHFtrans2011} and} \\ \textbf{Proposed Hybrid Scheme in Best Case}} & $(\frac{1}{2}M^{2}N+\frac{1}{3}M^{3})$\\\hline
\end{tabular}
\end{table}

This section compares the computational complexities and memory requirements of the existing and proposed V-BLAST algorithms by theoretical analysis. Note that, throughout this section, the computational cost of $j$ complex multiplications and $k$ complex additions is denoted by $(j,k)$. When $j=k$, the notation $(j,k)$ is abbreviated as $(j)$.



The dominant complexity of (\ref{derive3}) is $(\frac{1}{2}M^3)$, arising primarily from the computations of ${\bf{Q}}_{m-1}^{}{\bf{v}}_{m-1}$ and ${\bf{Q}}_{m-1}^{}+\omega_{m}^{}{\bf{\tilde w}}_{m-1}^{}{\bf{\tilde w}}_{m-1}^H$ for $m=2,3,\ldots,M$. In contrast, the complexity of (\ref{derive2Globecomm2005all3}) should be $(\frac{5}{6}M^3)$ rather than $(\frac{1}{2}M^3)$ as claimed in \cite{zhf6}. This complexity is mainly incurred by computing ${\bf{g}}_{m-1}={\bf{Q}}_{m-1}^{}{\bf{v}}_{m-1}^{}$, ${\bf{Q}}_{m-1}+{\bf{g}}_{m-1}{\bf{g}}_{m-1}^H/(\cdots)$, and ${\bf{\bar Q}}_{m-1}{\bf{v}}_{m-1}$. Therefore, compared with the inversion step in \cite{zhf5,zhf6} that computes the initial ${\bf{Q}}$ using (\ref{equQsmally26}) and (\ref{derive2Globecomm2005all3}), the proposed inversion step based on (\ref{equQsmally26}) and (\ref{derive3}) achieves a speedup of $(\frac{5}{6})/(\frac{1}{2})=1.67$ and requires only $\frac{1}{3}$ as many divisions.

The complexity reported in \cite{zhf6} for the above-mentioned inversion step used to compute the initial ${\bf Q}$ should be increased by $(\frac{5}{6}M^3-\frac{1}{2}M^3)=(\frac{1}{3}M^3)$. Then, the ``fastest known algorithm" prior to \cite{zhf6} and the recursive algorithm with speed advantage in \cite{zhf6} actually require the complexities of $(\frac{1}{2}M^2N +\frac{4}{3}M^3)$ and $(\frac{1}{2}M^2N+M^3)$, respectively, rather than $(\frac{1}{2}M^2N +M^3)$ and $(\frac{1}{2}M^2N +\frac{2}{3}M^3)$ as claimed in \cite{zhf6}. Notably, the latter complexity, i.e., $(\frac{1}{2}M^2N +\frac{2}{3}M^3)$, is exactly the dominant complexity of both the proposed recursive algorithm with speed advantage and that with both speed advantage and memory saving, since the computations of (\ref{equ:16}), (\ref{equ:BLAST_IC}), (\ref{equ:16Update}), and (\ref{dUpdate2388932sde3}) from Improvements \uppercase\expandafter{\romannumeral+2} and \uppercase\expandafter{\romannumeral+6} do not incur any dominant complexity. On the other hand, the recursive algorithm with memory saving in \cite{zhf6} requires $2M^2N +\frac{1}{6}M^3$ multiplications rather than $\frac{5}{2}M^2N +\frac{1}{6}M^3$ multiplications as stated in \cite[Eq. 27]{zhf6}. This is because it adopts Improvement \uppercase\expandafter{\romannumeral+3}, for which computing the initial ${\bf{Q}}$ using (\ref{QinitialOldXiaEqu5}) requires $\frac{3}{2}M^2N$ multiplications~\cite{zhf4}, instead of the $2M^2N$ multiplications claimed in \cite[Eq. 27]{zhf6}.

On the other hand, the worst-case, average-case, and best-case complexities\footnote{The recursive algorithm considered for comparison in \cite{refZHFtrans2011} has the same dominant complexity as the proposed recursive algorithms. Therefore, the comparison in \cite{refZHFtrans2011} can also be regarded as a comparison between the proposed recursive algorithms and the square-root algorithm.} of the square-root algorithm based on the inverse Cholesky factor are $(\frac{1}{2}M^{2}N+\frac{5}{6}M^{3}$, $\frac{1}{2}M^{2}N+\frac{1}{2}M^{3})$, $(\frac{1}{2}M^{2}N+\frac{2}{3}M^{3}$, $\frac{1}{2}M^{2}N+\frac{4}{9}M^{3})$, and $(\frac{1}{2}M^{2}N+\frac{M^{3}}{3})$, respectively~\cite{refZHFtrans2011}. In the best case, the proposed hybrid scheme reduces to the square-root algorithm in \cite{refZHFtrans2011} and therefore has the same complexity. The worst-case complexity of the proposed hybrid scheme is upper-bounded by the case in which switching to the square-root step provides no complexity reduction, such that ${\bf{q}}_{m}^H$ is computed using (\ref{ApdxJ21equ10bestJul7}) at every recursion. Computing (\ref{ApdxJ21equ10bestJul7}) for $m=M,M-1,\cdots,2$ is equivalent to first obtaining the initial ${\bf{Q}}_{M}$ from ${\bf{F}}_{M}$ using (\ref{ApdxJ21equ1}) and then recursively deflating the permuted ${\bf{Q}}_M$ through (\ref{equ:14}) for $m=M,M-1,\cdots,2$. The total dominant complexity of computing ${\bf{Q}}_{M}$ from ${\bf{F}}_{M}$ is equal to the complexity of $(\frac{M^{3}}{2})$ required to compute ${\bf{Q}}_{M}$ using (\ref{derive3}), because computing ${\bf{F}}_{M}$ and ${\bf{Q}}_{M}$ through (\ref{equ20_19}) and (\ref{ApdxJ21equ1}) requires complexities of $(\frac{M^{3}}{3})$~\cite{refZHFtrans2011} and $(\frac{M^{3}}{6})$~\cite{zhf_VTC08_6}, respectively. It therefore follows that the worst-case and average dominant complexities of the proposed hybrid scheme do not exceed those of either proposed recursive algorithm.

The dominant complexities of the presented V-BLAST algorithms are summarized in Table 1. To quantify the corresponding speedups, we assume $M=N$. As shown in Table 1, the actual speedup of the algorithm with speed advantage in \cite{zhf6} over the preceding ``fastest known algorithm" is $(\frac{11}{6})/(\frac{3}{2})=1.22$, rather than the value of $1.3$ reported in \cite{zhf6}. In comparison, either proposed recursive algorithm achieves speedups of $(\frac{3}{2})/(\frac{7}{6})=1.3$ and $(\frac{13}{6})/(\frac{7}{6})\approx 1.86$ over the algorithm with speed advantage and that with memory saving in \cite{zhf6}, respectively. Relative to either proposed recursive algorithm, the proposed hybrid scheme provides a best-case speedup of $(\frac{7}{6})/(\frac{5}{6})=1.4$, without increasing the dominant complexity in the worst-case or average-case scenario.


With respect to the ``fastest known algorithm" prior to \cite{zhf6},
 	the  algorithm with memory saving in \cite{zhf6} (i.e., \textbf{Algorithm 2})
 	costs more computations to 
 	save memories for storing ${\bf{R}}_{M}$,
 	and then only needs to store ${\bf{H}}_{M}$ and ${\bf{Q}}_{M}$. 
 	As a comparison,
 	the proposed recursive algorithm  with
 	both speed advantage and
 	memory  saving
 	only stores ${\bf{Q}}_{M}$ in the \textsl{recursion} phase, 
 	and only uses memories for storing ${\bf{H}}_{M}$ in the \textsl{initialization} phase since it 
 	 enables an in-place  implementation in which ${\bf{H}}_{M}$,  ${\bf{R}}_{M}$ and  ${\bf{Q}}_{M}$ sequentially occupy the same principal matrix-storage region.
 	Accordingly,
 	it can be concluded that
 	the proposed recursive algorithm reduces the principal matrix-storage requirement by approximately one half compared with the memory-saving algorithm in~\cite{zhf6}.\footnote{Only the memory required to store matrices of size at least $2\times2$ is considered.}

Before presenting the simulation results, we first specify the parameters $\beta$ for $m/M' \leq \beta$ and $\tau$ for $l_m / m > \tau$ in \textbf{Algorithm 4}. For a recursion $m$ initialized with $M' \neq m$, the theoretical complexity analysis detailed in the next paragraph compares the total complexity required to compute a square matrix ${\bf{F}}_m$ with the complexity reduction achieved by using ${\bf{F}}_m$ in the subsequent steps. This comparison indicates that ${\bf{F}}_m$ should be computed when

 \begin{equation}\label{mM35Equ329sd}
 	m/M \le 3/5.
 \end{equation}

Then solely through exhaustive search by simulations, we have found that, for a recursion $m$ initialized with $M'=m$, the proposed hybrid scheme should switch to the square-root step when

 \begin{equation}\label{lmm12djewdjsdss}
 	l_m  / m > 1/2.
 \end{equation}

%

We next detail the theoretical complexity analysis leading to (\ref{mM35Equ329sd}). At any recursion $m$ initialized with $M' \neq m$, ${\bf{F}}_m$ for the remaining $m$ undetected antennas can be obtained by first computing (\ref{equ:22}) and then iteratively applying (\ref{equ20_19}), which incurs a dominant complexity of $(\frac{m^3}{3})$, as discussed above. For simplicity, we consider only the dominant complexity reduction achieved by using the square matrix ${\bf{F}}_m$ to compute ${\bf{W}}_{{M'}-i}(i,:) {\bf{C}}_{{M'}-i} {\bf{W}}_{{M'}-i}^H$ ($i \leq m$) in (\ref{ApdxJ21equ10bestJul7}). At recursion $i$ ($i=m,m-1,\cdots,1$), the dominant complexity of this computation is $((m-i)i)$ when ${\bf{F}}_m$ is available and $((M'-i)i)$ when only ${\bf{F}}_{m \times {M'}}$ (${M'} > m$) is available. Therefore, the dominant complexity saved at recursion $i$ is $((M-i)i) - ((m-i)i) = ((M-m)i)$, and the corresponding total saving is $\sum\limits_{i = 1}^m ((M-m)i) \approx ((M-m)m^2 /2)$. To ensure that the complexity required to compute ${\bf{F}}_m$ does not exceed the resulting complexity reduction, $m$ should satisfy $m^3/3 \leq (M-m)m^2 /2$, which is equivalent to (\ref{mM35Equ329sd}).

\section{NUMERICAL RESULTS}

In this section, numerical experiments are conducted to compare the computational complexities of the presented V-BLAST algorithms. We set $N=M$ and vary the number of transmit and receive antennas from $2$ to $24$, with complexity measured by the total number of floating-point operations (flops). The benchmarks include the original recursive algorithm~\cite{zhf3}, the recursive algorithms in~\cite{zhf6}, and the square-root algorithm in~\cite{refZHFtrans2011}. Since the two proposed recursive algorithms have identical flop counts, only the one with both speed advantage and memory saving is shown and is referred to as the ``proposed recursive algorithm''. For \textbf{Algorithm 4}, $\tau=1/2$ and $\beta=3/5$ are used unless otherwise specified.

Fig. 1 compares the flop counts of the recursive algorithms and the proposed hybrid scheme. The numerical results are consistent with the analysis in Section VI. For $N=M$, the proposed recursive algorithm achieves speedups of approximately $1.3$ and $1.86$ over the recursive algorithms with speed advantage and memory saving in~\cite{zhf6}, respectively. The hybrid scheme substantially reduces the best-case complexity, while its worst-case complexity does not exceed that of the proposed recursive algorithm.

\begin{figure}
	\centerline{\includegraphics[width=3.5in]{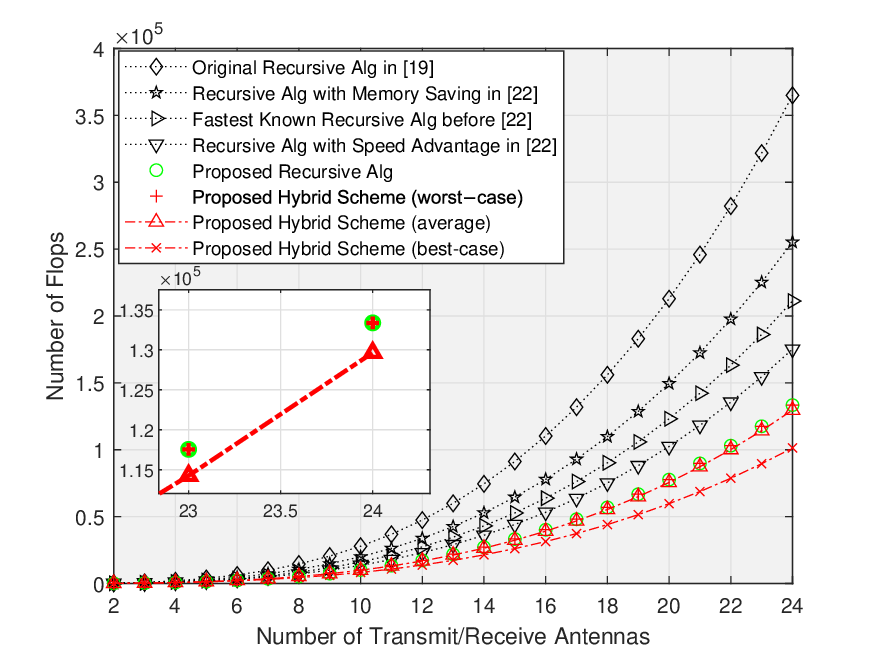}}
	\caption{Comparison of computational complexities among  the original recursive
		algorithm in \cite{zhf3},
		the recursive  algorithm with memory saving in \cite{zhf6},
		the  fastest known recursive algorithm before \cite{zhf6},
		the recursive algorithm with speed advantage in \cite{zhf6},
		the proposed recursive algorithm with
		both speed advantage and
		memory  saving, and the proposed hybrid scheme.\label{fig1}}
\end{figure}

Fig. 2 further compares the recursive algorithm with speed advantage in~\cite{zhf6}, the square-root algorithm in~\cite{refZHFtrans2011}, the proposed recursive algorithm, and the proposed hybrid scheme. In Fig. 2, “Proposed Recursive Alg (Inv. Chol. Impl.)” represents the inverse-Cholesky-factor-based implementation of the proposed recursive algorithm\footnote{It is the proposed hybrid scheme that never reduces the complexity by switching to the square root step. Accordingly, ${\bf{q}}_{m}^H$ is computed by (\ref{ApdxJ21equ10bestJul7}) at each recursion.}, which requires slightly more flops because of additional non-dominant operations for the introduction of the inverse Cholesky factor. Consistent with Fig. 1, the hybrid scheme approaches the best-case complexity of the square-root algorithm, preserves the worst-case advantage of the proposed recursive algorithm, and slightly reduces the average complexity. Moreover, nearly identical results are obtained for $\beta=1/2$ and $\beta=3/5$, indicating that the hybrid scheme is insensitive to moderate variations in $\beta$.

\begin{figure}
\centerline{\includegraphics[width=3.5in]{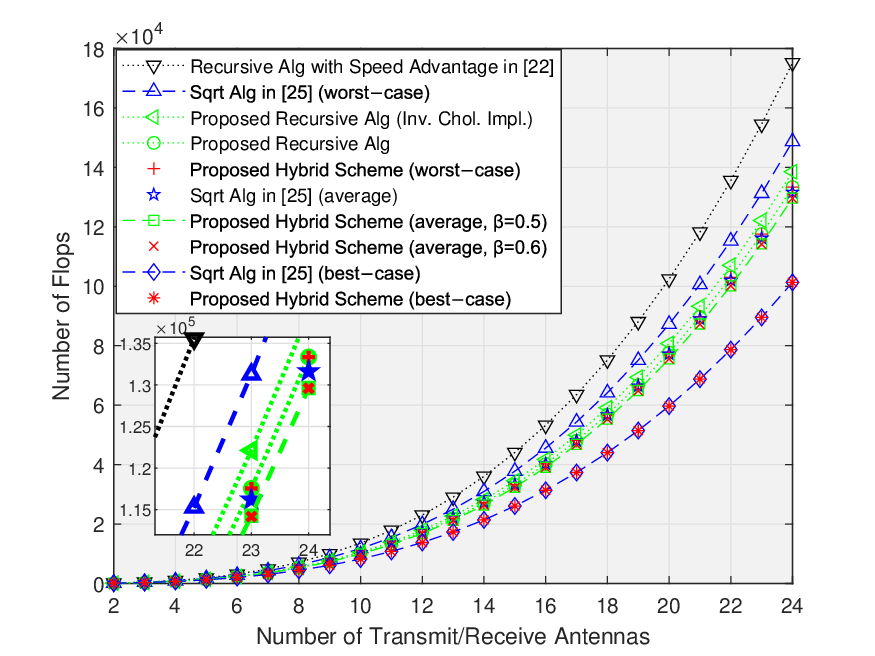}}
\caption{Comparison of computational complexities among  the recursive algorithm with speed advantage in \cite{zhf6},
	the square-root algorithm in \cite{refZHFtrans2011}, 
	the proposed recursive algorithm with
	both speed advantage and
	memory  saving, and the proposed hybrid scheme.\label{fig1}}
\end{figure}

\section{CONCLUSION}

This paper focuses on reducing the complexity and memory requirement of the recursive V-BLAST detection algorithms. Two improvements were proposed to reduce the cost of partitioned-matrix inversion and eliminate the dependence on the inverse covariance matrix during interference cancellation, respectively. The resulting recursive algorithms accelerate the matrix-inversion step by a factor of $1.67$ and achieve a speedup of $1.3$ over the existing algorithm with speed advantage. The proposed algorithm with both speed advantage and memory saving further achieves a speedup of approximately $1.86$ and reduces the principal matrix-storage requirement by about one half compared with the existing memory-saving algorithm. A hybrid scheme combining recursive and inverse-Cholesky-based square-root processing was also developed to reduce the best-case and worst-case complexities of the recursive and square-root algorithms, respectively, while slightly lowering the average complexity.

This work has focused on OSIC detection. Future research will extend the proposed recursive framework to ISIC detection by exploiting an equivalent-channel-based reordered representation, with the objective of further reducing its computational and memory requirements.

\section*{APPENDIX A. Derivation of (\ref{derive2})}

Equation (8) in \cite{InverseSumofMatrix8312}  can be written  as

\begin{equation}\label{Aorig239sd23}
	\left[ {\begin{array}{*{20}{c}}
			{\bf{A}}&{\bf{U}}\\
			{\bf{V}}&{\bf{D}}
	\end{array}} \right]^{-1}=\left[ {\begin{array}{*{20}{c}}
			{{\bf{\tilde A}}}&{{\bf{\tilde U}}}\\
			{{\bf{\tilde V}}}&{{\bf{\tilde D}}}
	\end{array}} \right]
\end{equation}%

with

\begin{subnumcases}{\label{PQRSWhole2309sd32}}
	{\bf{\tilde A}} = {{\bf{A}}^{ - 1}} \!+\! ({{\bf{A}}^{ - 1}}{\bf{U}}){({\bf{D}} \!-\! {\bf{V}}{{\bf{A}}^{ - 1}}{\bf{U}})^{ - 1}}({\bf{V}}{{\bf{A}}^{ - 1}}), &  \label{PQRSWhole2309sd32a}\\
	{\bf{\tilde U}} =  - ({{\bf{A}}^{ - 1}}{\bf{U}}){({\bf{D}} - {\bf{V}}{{\bf{A}}^{ - 1}}{\bf{U}})^{ - 1}},  &  \label{PQRSWhole2309sd32b}\\
	{\bf{\tilde V}} =  - {({\bf{D}} - {\bf{V}}{{\bf{A}}^{ - 1}}{\bf{U}})^{ - 1}}({\bf{V}}{{\bf{A}}^{ - 1}}),   &  \label{PQRSWhole2309sd32NoUse}\\
	{\bf{\tilde D}} = {({\bf{D}} - {\bf{V}}{{\bf{A}}^{ - 1}}{\bf{U}})^{ - 1}},  &  \label{PQRSWhole2309sd32c}
\end{subnumcases}%

which inverts a partitioned matrix.
By comparing (\ref{Aorig239sd23})
with (\ref{equ:11}) and (\ref{equQsmally26}),
we can conclude that
${\bf{A}}$, ${{\bf{A}}^{ - 1}}$,  ${\bf{U}}$, ${\bf{V}}$, ${\bf{D}}$,   ${\bf{\tilde A}}$, ${\bf{\tilde U}}$ and ${\bf{\tilde D}}$
can be replaced with  ${{\bf{R}}_{m - 1}}$,  ${{\bf{Q}}_{m - 1}}={\bf{R}}_{m-1}^{ - 1}$, ${{\bf{v}}_{m-1}}$, ${\bf{v}}_{m-1}^H$, ${\gamma _m}$,
${\bf{\bar Q}}_{m - 1}^{}$,
${{\bf{w}}_{m-1}^{} }$
and ${\omega _m}$,
respectively.
Accordingly,  we can
write
(\ref{PQRSWhole2309sd32a}),
(\ref{PQRSWhole2309sd32b}),
and
(\ref{PQRSWhole2309sd32c})
as

\begin{multline}\label{T2QQvmvQvvQ2390s2323}
	{\bf{\bar Q}}_{m - 1}^{} = {{\bf{Q}}_{m - 1}} + ({{\bf{Q}}_{m - 1}}{{\bf{v}}_{m-1}}) \\
	\times {({\gamma _m} - {\bf{v}}_{m-1}^H{{\bf{Q}}_{m - 1}}{{\bf{v}}_{m-1}})^{ - 1}}({\bf{v}}_{m-1}^H{{\bf{Q}}_{m - 1}}),
\end{multline}
\begin{equation}\label{w2QvgvQv2334df342}
	{{\bf{w}}_{m-1}^{} } \!\!=\!\!  - ({{\bf{Q}}_{m - 1}}{{\bf{v}}_{m-1}}){({\gamma _m} \!-\! {\bf{v}}_{m-1}^H{{\bf{Q}}_{m - 1}}{{\bf{v}}_{m-1}})^{ - 1}},
\end{equation}%

and (\ref{derive2b}),  respectively. Finally,  we substitute (\ref{derive2b}) into (\ref{w2QvgvQv2334df342})
to obtain
(\ref{derive2a}),  and substitute (\ref{derive2b}) and (\ref{derive2a}) into (\ref{T2QQvmvQvvQ2390s2323})
to obtain
(\ref{derive2c}).

\section*{APPENDIX B. Derivation of (\ref{s2t239032sd23s}) and (\ref{t2tstq349sd23ds23})}

It follows from (\ref{d2Qzt934dscerwsdf34}) that

\begin{subnumcases}{\label{}}
	{{\mathbf{\bar{d}}}_{m}}={{\mathbf{Q}}_{m}}(1:m-1,:){{\mathbf{z}}_{M}}(1:m)-{{\mathbf{\bar t}}_{m}}, &  \label{d2Qzt9032sd23sdt34}\\
	{{\mathbf{d}}_{m}}(m)=\mathbf{q}_{m}^{H}{{\mathbf{z}}_{M}}(1:m)-{{\mathbf{t}}_{m}}(m).  &  \label{d2qzt92332dscrfd}
\end{subnumcases}%

Moreover, it can be observed from (\ref{equQsmally26}) that

\begin{subnumcases}{\label{}}
{{\mathbf{Q}}_{m}}(1:m-1,:)=\left[ \begin{matrix}
\mathbf{\bar{Q}}_{m-1}^{{}} & \mathbf{\bar{q}}_{m}^{{}}  \\
\end{matrix} \right], &  \label{Q2Qq9239sd3sd}\\
\mathbf{q}_{m}^{H}=\left[ \begin{matrix}
\mathbf{\bar{q}}_{m}^{H} & {{\omega }_{m}}  \\
\end{matrix} \right].  &  \label{q2qOmega3sd32ds}
\end{subnumcases}%

Substituting (\ref{derive2c}) into (\ref{Q2Qq9239sd3sd}) yields
${{\mathbf{Q}}_{m}}(1:m-1,:)  =\left[
{{\bf{Q}}_{m - 1}^{} + \omega _m^{ - 1}{\bf{\bar q}}_m^{}{\bf{\bar q}}_m^H} \quad {{\bf{\bar q}}_m^{}}\right]$,
which is substituted into (\ref{d2Qzt9032sd23sdt34})  to obtain

\begin{align}
		 {{\bf{\bar d}}_m} &= \left[
		{{\bf{Q}}_{m - 1}^{} + \omega _m^{ - 1}{\bf{\bar q}}_m^{}{\bf{\bar q}}_m^H} \quad {{\bf{\bar q}}_m^{}}\right] {{\bf{z}}_M}(1:m) - {{\bf{\bar t}}_m}  \notag \\
		&= ({\bf{Q}}_{m - 1}^{} + \omega _m^{ - 1}{\bf{\bar q}}_m^{}{\bf{\bar q}}_m^H){{\bf{z}}_M}(1:m - 1) \notag\\
& + {{\bf{z}}_M}(m){\bf{\bar q}}_m^{} - {{\bf{\bar t}}_m}.   \label{d2Qzt39sd23sd5f}
\end{align}%

On the other hand, substituting (\ref{q2qOmega3sd32ds}) into (\ref{d2qzt92332dscrfd}) yields

\begin{align}
{{\bf{d}}_m}(m)&= \left[ {\begin{array}{*{20}{c}}
{{\bf{\bar q}}_m^H}&{{\omega _m}}
\end{array}} \right]{{\bf{z}}_M}(1:m) - {{\bf{t}}_m}(m)  \notag \\
&= {\bf{\bar q}}_m^H{{\bf{z}}_M}(1:m - 1) + {\omega _m}{{\bf{z}}_M}(m) - {{\bf{t}}_m}(m).  \label{d2qzwzt932sdcx43sd}
\end{align}%

Equation (\ref{s2t239032sd23s}) follows directly by substituting (\ref{d2qzt92332dscrfd}) into (\ref{equ:16Update}), which gives ${{\hat s}_{{p_m}}} = {\bf{q}}_m^H{{\bf{z}}_M}(1:m) - \left( {{\bf{q}}_m^H{{\bf{z}}_M}(1:m) - {{\bf{t}}_m}(m)} \right)$. To derive (\ref{t2tstq349sd23ds23}), we substitute (\ref{d2Qzt934dscerwsdf34}) (with $n=n-1$), (\ref{d2Qzt39sd23sd5f}) and (\ref{d2qzwzt932sdcx43sd}) into (\ref{dUpdate2388932sde3}), obtaining

									\begin{small}
										\begin{gather*}
											{{\bf{Q}}_{m - 1}}{{\bf{z}}_M}(1:m - 1) - {{\bf{t}}_{m - 1}} = ({\bf{Q}}_{m - 1}^{} + \omega _m^{ - 1}{\bf{\bar q}}_m^{}{\bf{\bar q}}_m^H) \times \\
											{{\bf{z}}_M}(1:m - 1) + {{\bf{z}}_M}(m){\bf{\bar q}}_m^{} - {{\bf{\bar t}}_m} - \\
											\left( {{s_{{p_m}}} + \left( {{\bf{\bar q}}_m^H{{\bf{z}}_M}(1:m - 1) + {\omega _m}{{\bf{z}}_M}(m) - {{\bf{t}}_m}(m)} \right)} \right){{\bf{\bar q}}_m}/{\omega _m},
										\end{gather*}
									\end{small}%

									which can be directly simplified to (\ref{t2tstq349sd23ds23}).

\section*{APPENDIX C. Verification of (\ref{ApdxJ21equ7})}

In this appendix, we verify (\ref{ApdxJ21equ7}) for $m=M,M-1,\cdots,1$ by induction. For $m=M$, (\ref{ApdxJ21equ7}) follows directly from (\ref{ApdxJ21equ1}) by setting ${\bf{W}}_{0}$ and ${\bf{C}}_{0}$ to empty matrices. Throughout the following derivation, the rows and columns of ${\bf{Q}}_{m}$, together with the rows of ${\bf{F}}_{m \times M}$ and ${\bf{W}}_{M-m}$, are assumed to have been appropriately permuted according to the optimal detection order.

To obtain (\ref{ApdxJ21equ7}) for $m=M-1$ from that for $m=M$, we substitute (\ref{equQsmally26}) with $m=M$ into the latter, yielding ${{\bf{\bar Q}} _{M - 1}^{} }={\bf{F}}_{(M-1)\times M}{\bf{F}}_{(M-1)\times M}^H$ with ${\bf{F}}_{(M-1)\times M}$ consisting of the first $(M-1)$ rows of ${\bf{F}}_M$. Substituting this result into (\ref{equ:14}) with $m=M$ establishes (\ref{ApdxJ21equ7}) for $m=M-1$.

Similarly, to derive (\ref{ApdxJ21equ7}) for $m=m-1$ from that for any $m \leq M-1$, substituting (\ref{equQsmally26}) into (\ref{ApdxJ21equ7}) gives $	{\bf{\bar Q}}_{m-1}  = {\bf{F}}_{(m-1)\times M} {\bf{F}}_{(m-1) \times M}^H  - {\bf{\bar W}}_{M-m} {\bf{C}}_{M-m} {\bf{\bar W}}_{M-m}^H$. Further substituting this expression into (\ref{equ:14}) establishes (\ref{ApdxJ21equ7}) for $m=m-1$. Therefore, (\ref{ApdxJ21equ7}) holds for all $m=M,M-1,\cdots,1$.

\end{document}